\documentclass[prl,twocolumn,superscriptaddress,floatfix,showpacs,amsmath,longbibliography,nofootinbib,amssymb]{revtex4-1}
\usepackage[colorlinks=true,citecolor=blue,linkcolor=blue]{hyperref}
\usepackage[usenames]{color}
\usepackage{subfigure}
\usepackage{dcolumn}
\usepackage{mathrsfs}
\usepackage{bm}
\usepackage{amsmath,amssymb,epsfig,float,graphics}

\usepackage{graphicx}
\usepackage{dcolumn}
\usepackage{bm}
\usepackage{hyperref}
\usepackage{xcolor}
\usepackage{amsmath, amssymb, amsfonts}
\usepackage{mathtools}
\usepackage{subfigure}
\usepackage{mathrsfs}

\newcommand{\colb}[1]{\textcolor{blue}{#1}}

\definecolor{purple1}{rgb}{128,0,128}

\newcommand{\bea}{\begin{eqnarray}}
\newcommand{\ea}{\end{eqnarray}}

\DeclareMathOperator{\sech}{sech}

\begin{document}
\baselineskip=0.45 cm

\title{Verifying the upper bound on the speed of 
scrambling with the analogue Hawking radiation of trapped ions}

\author{Zehua Tian}
\email{tianzh@ustc.edu.cn}
\affiliation{Hefei National Laboratory for Physical Sciences at the Microscale and Department of Modern Physics, University of Science and Technology of China, Hefei 230026, China}
\affiliation{CAS Key Laboratory of Microscale Magnetic Resonance, University of Science and Technology of China, Hefei 230026, China}
\affiliation{Synergetic Innovation Center of Quantum Information and Quantum Physics, University of Science and Technology of China, Hefei 230026, China}

\author{Yiheng Lin}
\email{Yiheng@ustc.edu.cn}
\affiliation{Hefei National Laboratory for Physical Sciences at the Microscale and Department of Modern Physics, University of Science and Technology of China, Hefei 230026, China}
\affiliation{CAS Key Laboratory of Microscale Magnetic Resonance, University of Science and Technology of China, Hefei 230026, China}
\affiliation{Synergetic Innovation Center of Quantum Information and Quantum Physics, University of Science and Technology of China, Hefei 230026, China}

\author{Uwe R. Fischer}
\affiliation{Seoul National University, Department of Physics and Astronomy, Center for Theoretical Physics, Seoul 08826, Korea}


\author{Jiangfeng Du}
\email{djf@ustc.edu.cn}
\affiliation{Hefei National Laboratory for Physical Sciences at the Microscale and Department of Modern Physics, University of Science and Technology of China, Hefei 230026, China}
\affiliation{CAS Key Laboratory of Microscale Magnetic Resonance, University of Science and Technology of China, Hefei 230026, China}
\affiliation{Synergetic Innovation Center of Quantum Information and Quantum Physics, University of Science and Technology of China, Hefei 230026, China}

\begin{abstract}
A general bound on the Lyapunov exponent of a 
quantum system is given by 
$\lambda_L\leq2\pi\,T/\hbar$, where $T$  is the system temperature, as established by 
Maldacena, Shenker, and Stanford (MSS). This upper bound is saturated when the system under consideration is the exact holographic dual of a black hole. 
It has also been shown that an inverted harmonic oscillator (IHO) may exhibit the 
behavior of thermal energy emission, in close analogy to the Hawking radiation emitted by black holes.  
We demonstrate that the Lyapunov exponent of the IHO indeed saturates the MSS bound, 
with an effective temperature equal to the analogue black hole radiation temperature, 
and propose using a trapped ion as a physical implementation of the IHO. 
We    
derive the corresponding out-of-time-ordered correlation function 
(OTOC) diagnosing quantum chaos, 
and theoretically show, for an experimentally realizable setup, 
that the effective temperature of the trapped-ion-IHO matches the upper 
MSS bound for the speed of scrambling. 
\end{abstract}

\baselineskip=0.45 cm
\maketitle
\newpage
The Hawking effect \cite{Hawking1, Hawking1975}, describing particle creation by quantum vacuum fluctuations in the presence of a black hole event horizon, can be mimicked in the laboratory, as argued by Unruh \cite{PhysRevLett.46.1351}.  
Since his seminal work, and with increasing experimental capabilities, a veritable explosion of 
interest occurred in what was dubbed {\em analogue gravity} \cite{analogue-gravity},    
involving, for example, Bose-Einstein condensates \cite{PhysRevLett.85.4643, BLV,PhysRevLett.91.240407, PhysRevA.69.033602,Ralf,PhysRevLett.115.025301,Carusotto_2008,BEC1, BEC2, BEC3, Renaud,PhysRevLett.124.060401,PhysRevLett.118.130404,Tyler} light in nonlinear media \cite{Philbin1367, PhysRevLett.105.203901, PhysRevLett.122.010404}, magnets \cite{PhysRevLett.118.061301,PhysRevLett.122.037203}, or superconducting circuits \cite{PhysRevLett.103.087004, PhysRevD.94.081501, PhysRevD.95.125003,PhysRevD.100.065003,superconducting-circuit}. 

Gauge-gravity duality \cite{GGD1, PhysRevD.58.046004} represents a powerful 
tool enabling the description of quantum gravity by certain classes of  
large-N gauge theories. 
Recently, it was conjectured that the Lyapunov exponent $\lambda_L$
characterizing the rate of growth of chaos in thermal quantum many-body systems 
has a rigorous bound, $\lambda_L\leq2\pi\,T/\hbar$ \cite{bound}. 
This bound is saturated when the quantum system 
is exactly holographic dual to a black hole \cite{QBC1, Sekino_2008, Franz:2018cqi}, 
and thus may capture an essential feature of the gauge-gravity duality. 
Conversely, the bound might suggest that a 
quantum system with given $\lambda_L$ has a minimal temperature $T\ge\hbar\lambda_L/2\pi$. 
Thus a system with $\lambda_L\neq 0$, although possessing zero temperature classically, 
may exhibit thermality in the semiclassical regime \cite{PhysRevLett.122.101603}, that is
$\lambda_L\neq 0, T\neq 0$, when $\hbar\neq 0$.
This recalls a salient property of black holes: While being completely black classically, 
they semiclassically radiate at the Hawking temperature.  

The quantum dynamics of the IHO exhibits thermal behavior with a temperature depending
on the 
Lyapunov exponent $\lambda_L$ \cite{PhysRevLett.122.101603, Betzios, Maldacena_2005}. 
The IHO model has, therefore, been used to 
explore the MSS bound on system temperature from a different angle,  
revealing its relation to analogue Hawking radiation 
 \cite{PhysRevLett.122.101603},
and to study the scattering outside a black hole horizon quantum mechanically \cite{Betzios, Maldacena_2005}. 
However,  to experimentally verify whether the claimed duality between black holes and the fastest quantum scramblers  
indeed exists in nature, and how to extract the Lyapunov exponent within an experimentally accessible system 
are still open issues. 
To fill this gap, we propose below an implementation of the IHO with a trapped ion.
Specifically, we propose to determine the Lyapunov exponent by measuring the IHO's 
OTOC \cite{1969JETP1200L, bound}. 
To establish gauge-gravity duality, we show that IHO analogue Hawking temperature and  
Lyapunov exponent indeed fulfill $\lambda_L= 2\pi\,T/\hbar$. 

Trapped ions, featuring a unique level of fidelity in preparation, control, and readout of quantum states, have been extensively used to simulate relativistic quantum physics, 
such as Zitterbewegung \cite{PhysRevLett.98.253005, ZMD, PhysRevLett.120.160403}, the 
Klein paradox \cite{PhysRevLett.106.060503, PhysRevA.82.020101}, and (analogue) cosmological particle creation  
\cite{PhysRevLett.94.220401, PhysRevLett.99.201301, PhysRevLett.123.180502, Menicucci_2010}. 
We demonstrate that our proposal furnishes an implementation of exact 
gauge-gravity duality within current experimental reach for trapped ions. 

The one-dimensional IHO, with Hamiltonian ($\alpha >0$)
 \begin{eqnarray}\label{Hamiltonian}
H=\frac{{p}^2}{2m}-\frac{\alpha}{2}{x}^2, 
\end{eqnarray}
yields the classical trajectory 
$x(t)=c_1\,e^{\sqrt{\alpha/m}t}+c_2\,e^{-\sqrt{\alpha/m}t}$ (see discussion below for the relation of the  IHO ``particle" to the actual ion).
This deterministic evolution is exponentially sensitive to the initial condition specified by $c_1$ and $c_2$, 
and the Lyapunov exponent $\lambda_L=\sqrt{\alpha/m}$.  
In the quantum version of the IHO, the OTOC 
for momentum and position operators is given by: 
\begin{eqnarray}\label{OTOC}
C(t)=-\langle[\hat{x}(t),\hat{p}(0)]^2\rangle\sim\hbar^2e^{2\lambda_Lt}. 
\end{eqnarray}
Applying the Lyapunov exponent bound  \cite{bound}, $\lambda_L\leq2\pi\,T/\hbar$, to the 
quantum IHO predicts the existence of a lower bound on temperature 
$T\geq\hbar\lambda_L/2\pi$, as conjectured in \cite{PhysRevLett.122.101603}. 
For the classical IHO ($\hbar\rightarrow 0$), which is both nonthermal and deterministic, 
this inequality becomes an equality as $T\coloneqq 0$.  
Conversely, in the semiclassical regime, following \cite{PhysRevLett.122.101603},  
the classical Lyapunov exponent $\lambda_L$ acquires a 
quantum correction ${\cal\,O}(\hbar)$, so that the right-hand side of the inequality 
becomes $T\geq\frac{\hbar}{2\pi}[\lambda_L+{\cal\,O}(\hbar)]=\frac{\hbar}{2\pi}\lambda_L+{\cal\,O}(\hbar^2)$. 
This suggests that, at least on the semiclassical level, an effective temperature 
of ${\cal\,O}(\hbar)$ is induced, which closely resembles the situation of black hole thermodynamics: While black holes are classical solutions of general relativity, semiclassically, they create a thermal bath of radiation 
 \cite{Hawking1, Hawking1975}. 

To establish the correspondence of black hole radiation and quantum behavior 
of the IHO, we consider scattering off the IHO potential. 
Defining the light-cone-type 
operators $\hat{u}^\pm=(\hat{p}^\prime\pm\hat{x}^\prime)/\sqrt{2}$ with $\hat{p}^\prime=\hat{p}/\sqrt{m}$ and $\hat{x}^\prime=\sqrt{\alpha}\hat{x}$, we can represent 
\eqref{Hamiltonian} in the $u^\pm$ basis as $H_\pm=\mp\,i\hbar\lambda_L\big(u^\pm\partial_{u^\pm}+\frac{1}{2}\big)$, respectively. 
The in- and outgoing energy eigenstates 
are of the form $\frac{1}{\sqrt{2\pi\hbar\lambda_L}}(\pm\,u^\pm)^{\pm\,i\frac{\varepsilon}{\hbar\lambda_L}-\frac{1}{2}}\Theta(\pm\,u^\pm)$, where $\Theta(u^\pm)$ is Heaviside step function, and $\varepsilon$ is energy. 
Incoming and outgoing
states are connected by a Mellin transform, which gives the scattering matrix ($S$ matrix) 
in the form \cite{Betzios, Maldacena_2005} 
\begin{eqnarray}\label{S-Matrix}
\nonumber
S&=&\frac{1}{\sqrt{2\pi}}\exp\bigg[-i\frac{\varepsilon}{\hbar\lambda_L}\log\hbar\lambda_L\bigg]\Gamma\bigg(\frac{1}{2}-i\frac{\varepsilon}{\hbar\lambda_L}\bigg)
\\    \nonumber
&&\times
\left(
\begin{array}{cccc}
e^{-i\frac{\pi}{4}}e^{-\frac{\pi\,\varepsilon}{2\hbar\lambda_L}} & e^{i\frac{\pi}{4}}e^{\frac{\pi\,\varepsilon}{2\hbar\lambda_L}}
\\
e^{i\frac{\pi}{4}}e^{\frac{\pi\,\varepsilon}{2\hbar\lambda_L}} & e^{-i\frac{\pi}{4}}e^{-\frac{\pi\,\varepsilon}{2\hbar\lambda_L}}
\end{array}
\right), \label{Smatrix} 
\end{eqnarray}
where $\Gamma$ is the Gamma function. From the $S$ matrix, the transmission and reflection probability are respectively given by, using 
$|\Gamma\big(\frac{1}{2}-i\frac{\varepsilon}{\hbar\lambda_L}\big)|^2=\pi\sech(\frac{\pi\varepsilon}{\hbar\lambda_L})$, 
\begin{eqnarray}\label{TR}
|{\mathcal T}(\varepsilon)|^2=\frac{1}{1+e^{2\pi\,\varepsilon/\hbar\lambda_L}},~~|{\mathcal R}(\varepsilon)|^2=\frac{1}{1+e^{-2\pi\,\varepsilon/\hbar\lambda_L}}, \label{TandR}
\end{eqnarray}
corresponding to a thermal distribution, with a temperature given by $T=\hbar\lambda_L/2\pi$. 

\begin{figure}[t]
\centering
\includegraphics[width=0.45\textwidth]{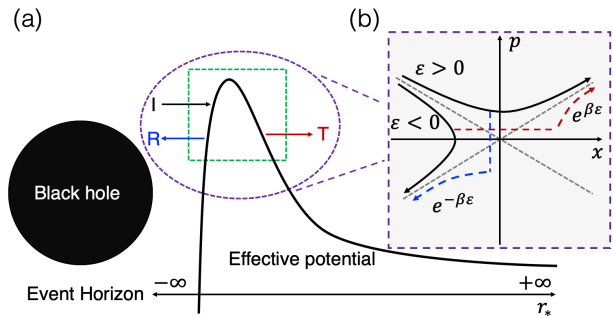}
\caption{(Color online) (a) Scattering off an effective potential approximately equal to an inverse harmonic potential, outside a black hole, where $r_\ast$ is the ``tortoise coordinate", and the event horizon is located at $r_\ast\rightarrow-\infty$. 
(b) Scattering via the inverse harmonic potential leads to classical trajectories of incoming particles  (solid lines) and particle tunneling (broken lines) near the hyperbolic fixed point $(x, p)=(0, 0)$ in phase space. The dotted lines are the separatrices $(\varepsilon=0)$, and $\beta=1/T$. }
\label{fig1}
\end{figure}

As illustrated in Fig. \ref{fig1}, we can understand the transmission coefficient above 
by the quantum mechanical tunneling process through a barrier 
of the particle with negative energy $\varepsilon$ moving towards the inverted harmonic potential from the left $(x\to0)$, with a tunneling probability $|\mathcal{T}(-\varepsilon)|^2$.  
Similarly, the incoming positive energy particle $(\varepsilon>0)$ from the right
may be reflected by the potential quantum-mechanically, with probability $|{\mathcal R}(-\varepsilon)|^2$. 

A relativistic particle, when close to a black hole horizon (pulled by external  
scalar or electromagnetic forces so that it does not fall into the black hole),  
possesses an effective Lagrangian corresponding to the 
Hamiltonian \eqref{Hamiltonian} \cite{PhysRevD.95.024007}. 
The maximal Lyapunov exponent of the particle's motion is  
found to be the surface gravity of the black hole when the total potential presents unstable maximum \cite{PhysRevD.95.024007}. The Lyapunov exponent is then given by 
$\lambda_L=\kappa=\frac{1}{2}\sqrt{g^\prime(r_0)f^\prime(r_0)}$, for the metric $ds^2=-f(r)dt^2+dr^2/g(r)+r^2(d\theta^2+\sin^2\theta\,d\phi^2)$. Here $g^\prime(r_0)$ and $f^\prime(r_0)$ are derivatives with respect to $r$ evaluated at the horizon $r=r_0$. 

Numerous variants of (re-)deriving Hawking's result exist  
\cite{Hawking1, Hawking1975, Gibbons, Parikh, Robinson, Satoshi, Peet:2000hn, STROMINGER199699, PhysRevD.77.024031}, ranging, e.g., from Hawking calculating the Bogoliubov coefficients between the quantum scalar field modes of the ``in" and ``out" vacuum states \cite{Hawking1, Hawking1975}, to an open quantum system approach \cite{PhysRevD.77.024031}. Here, we 
follow Parikh and Wilzcek \cite{Parikh}, employing the tunnelling probability through a classically forbidden region.

Consider our particle tunneling through an effective black hole potential due to a general spherically symmetric  metric $ds^2=-f(r)dt^2+dr^2/g(r)+r^2(d\theta^2+\sin^2\theta\,d\phi^2)$. 
On the semiclassical level under consideration here, we can employ the 
WKB approximation to calculate the tunnelling (transmission) probability through the horizon. 
To leading order in $\hbar$, it is given by \cite{Kerner2008}
\begin{eqnarray}
\Gamma=\exp\bigg[-\frac{4\pi}{\sqrt{g^\prime(r_0)f^\prime(r_0)}}\varepsilon\bigg]\coloneqq\exp[-\varepsilon/T]. 
\end{eqnarray}
The above tunneling rate implies Hawking radiation with 
temperature $T=\sqrt{g^\prime(r_0)f^\prime(r_0)}/4\pi$.
We conclude that IHO and black hole thus share the 
property that while classically not behaving as if being in thermal equilibrium with a bath, thermality is acquired when 
quantum mechanics is taken into account.  
This formal analogy forms the basis of our proposal to observe analogue 
Hawking radiation in an experimentally feasible  quantum IHO system, to which we now proceed.

We propose using a trapped ion to experimentally implement the IHO.
In our scenario, we monitor one of the oscillation directions of the ion,  
with associated momentum and position operators in the lab frame denoted as $\hat{P}$ and $\hat{X}$, respectively.
In particular, the ion's oscillation frequency $\omega_0$ can be 
periodically modulated by an applied voltage on the trap electrodes \cite{RevModPhys.75.281}.
The corresponding Hamiltonian is given by $H=\frac{\hat{P}^2}{2M}+\frac{1}{2}M\omega^2(t)\hat{X}^2$,
where $\omega(t)=\omega_0[1-\xi\cos(\omega_mt+\phi)]$ denotes the time-dependent frequency, $\xi$ is the modulation depth, $\phi$ is the initial phase of the modulation, and $M$ is the mass of the ion. 
Imposing $0<\xi\ll1$, and $\omega_m=2\omega_0$, 
the Hamiltonian, in rotating-wave approximation, 
can be rewritten in the form of the IHO 
in Eq.~\eqref{Hamiltonian}, with $m=2M/\xi$, and $\alpha=\frac{1}{4}m\omega_0^2\xi^2=m\omega^2$. The momentum and position 
operators have been redefined by $\hat{p}=-i\sqrt{\frac{m\omega\hbar}{2}}(\hat{a}e^{i\phi/2}-\hat{a}^\dagger\,e^{-i\phi/2})$,
and $\hat{x}=\sqrt{\frac{\hbar}{2m\omega}}(\hat{a}e^{i\phi/2}+\hat{a}^\dagger\,e^{-i\phi/2})$, respectively. Note that the phase $\phi$ is tunable by appropriately choosing the initial phase of driving.

To diagnose the \colb{quantum} chaotic motion of the IHO, we propose to measure its OTOC using the 
protocol proposed in Ref.~\cite{blocher2020measuring} 
which does not require the reversal of time evolution. Here, 
the OTOC is defined by  $C(t)=\mathrm{Tr}[\rho_0W^\dagger(t)V^\dagger\,W(t)V]$, 
where $W(t)=U^\dagger(t, 0)WU(t, 0)$ and $V=V(0)$ are operators evaluated in the
Heisenberg picture at times $0$ and $t$, respectively \cite{1969JETP1200L, bound}. 

Assuming the operator $V$
to be a projection operator onto an initially pure state, $V=\rho_0$  \cite{blocher2020measuring},  the OTOC can be reduced to $C(t)=|\langle\,W(t)\rangle|^2$.
To measure the OTOC, we therefore need 
to prepare a pure initial state for the degree of freedom of the ion's motion. For our 
experimental proposal, we use for concreteness a single $^9\mathrm{Be}^+$ ion trapped in a strong radio-frequency Paul trap with a pseudopotential trap frequency 
of $\omega_0/2\pi=10\,\mathrm{MHz}$. The motional ground state can be prepared 
by a two-stage laser cooling process. By Doppler cooling, all the motional modes of the ion could be cooled down to near the Doppler limit (with an average phonon number $\bar{n}$ of typically a few to ten quanta), through driving the 
$^2S_{1/2}$ to $^2P_{3/2}$ dipole transition. 
We can further cool the motional mode of interest to its ground state using sideband cooling with stimulated Raman transitions between the ion's two hyperfine ground states, i.e., between $^2S_{1/2}(F=2, m_F=2)$ and $^2S_{1/2}(F=1, m_F=1)$, denoted by $|\downarrow\rangle$ and $|\uparrow\rangle$, respectively, and are separated by $\sim1.25\,\mathrm{GHz}$ \cite{PhysRevA.42.2977, PhysRevLett.75.4011}. 
The spin initialization and ground state cooling allow us to prepare the ion in the 
$|S=\downarrow, n=0\rangle$ state eventually. By applying appropriate laser pulse sequences on blue and red sidebands or the carrier, 
we can create  arbitrary quantum superpositions of Fock states \cite{PhysRevA.55.1683, PhysRevA.57.2096, PhysRevLett.90.037902, PhysRevLett.92.113004}. 

\begin{figure}[t]
\centering
\subfigure[]{\includegraphics[width=0.2\textwidth]{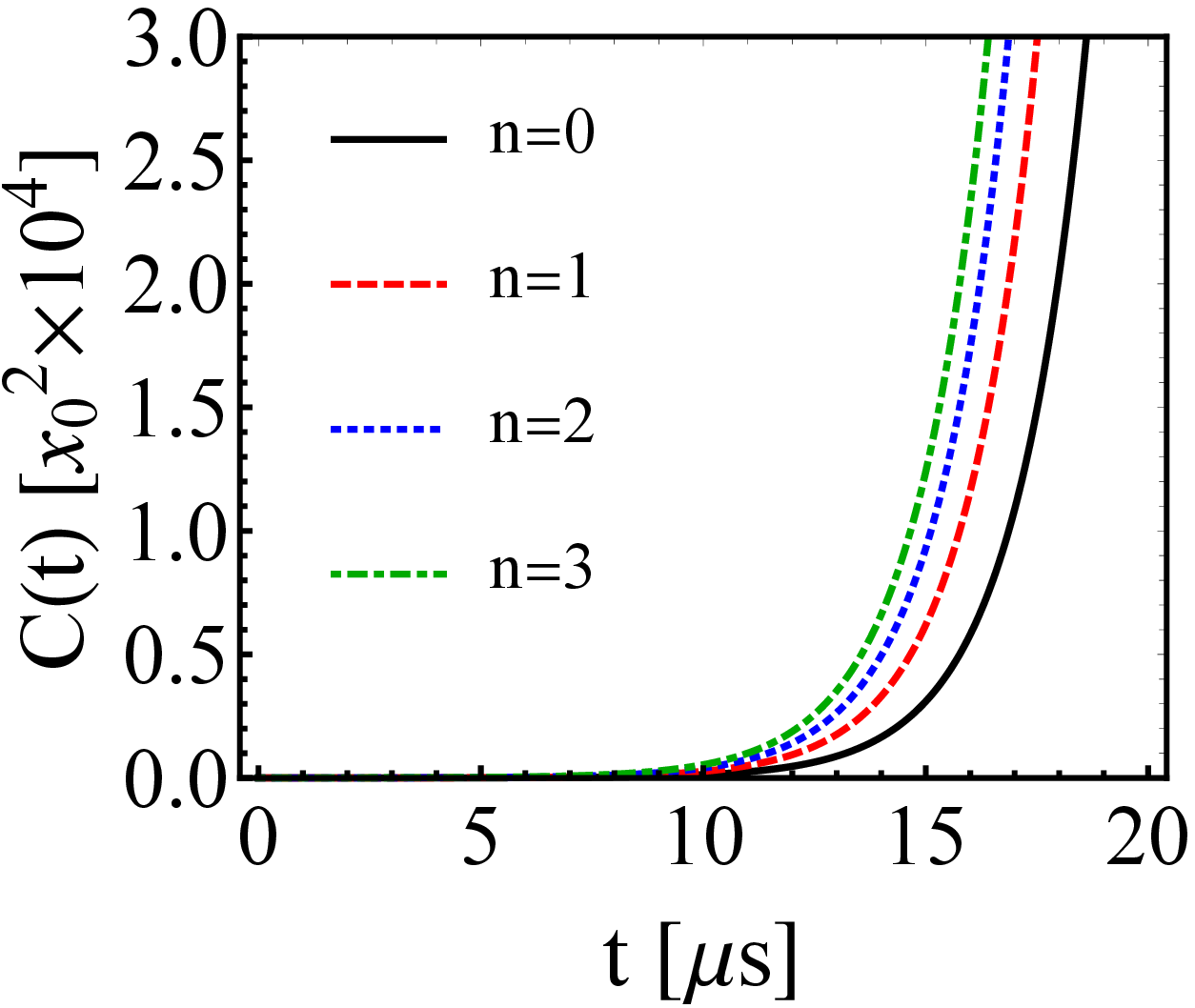}}\hspace*{2em}
\subfigure[]{\includegraphics[width=0.2\textwidth]{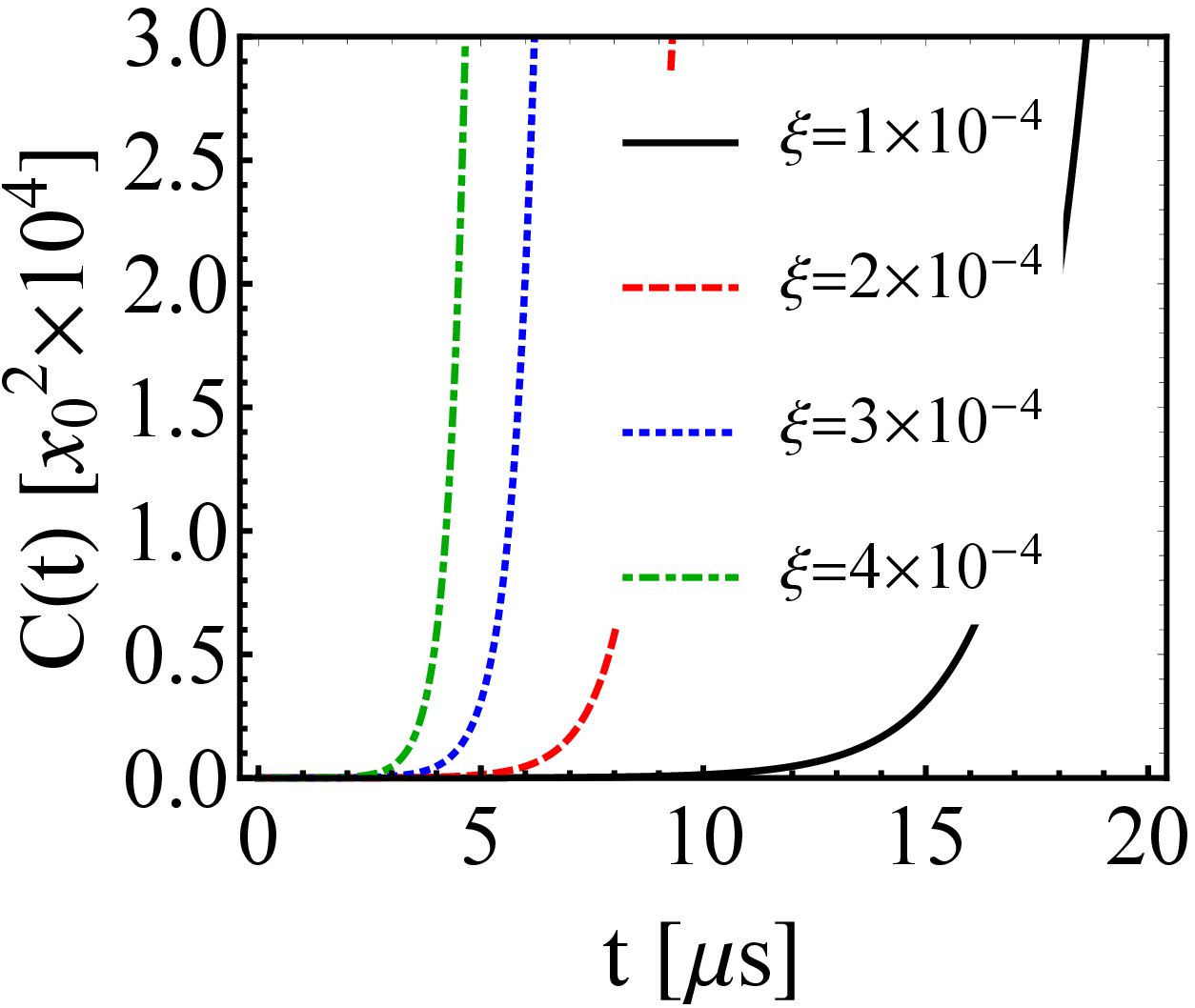}}
\subfigure[]{\includegraphics[width=0.2\textwidth]{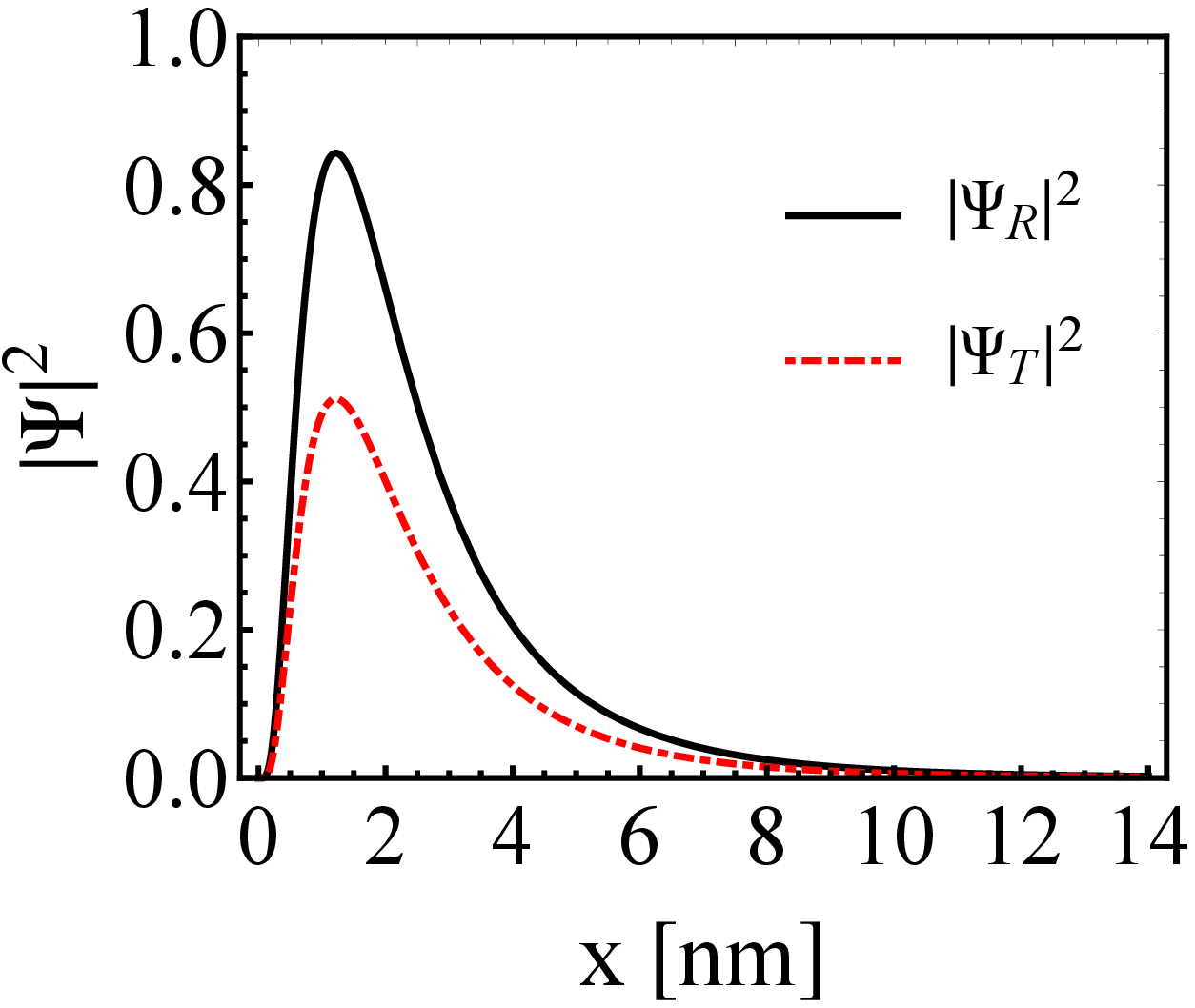}}\hspace*{2em}
\subfigure[]{\includegraphics[width=0.2\textwidth]{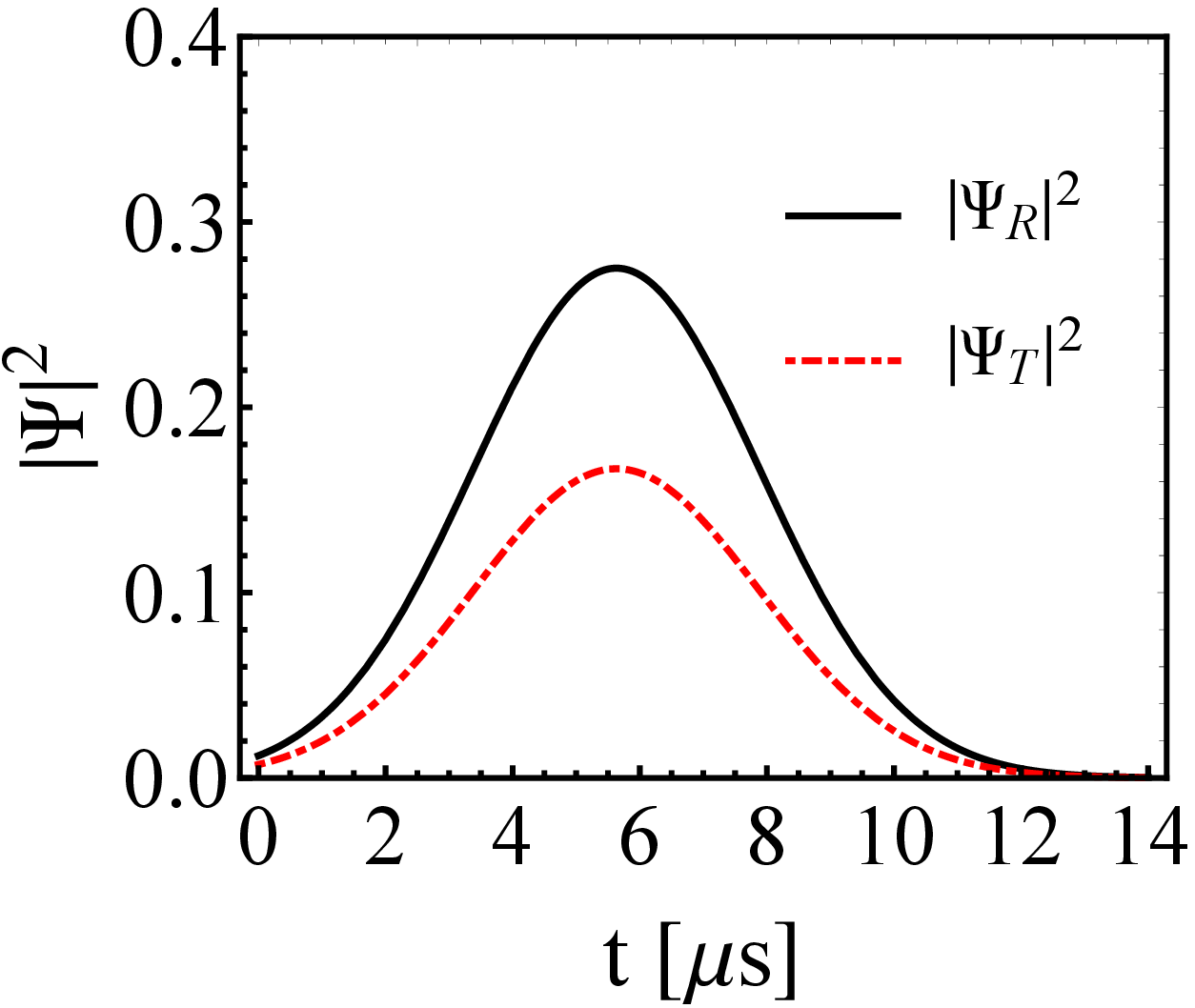}}
\subfigure[]{\includegraphics[width=0.25\textwidth]{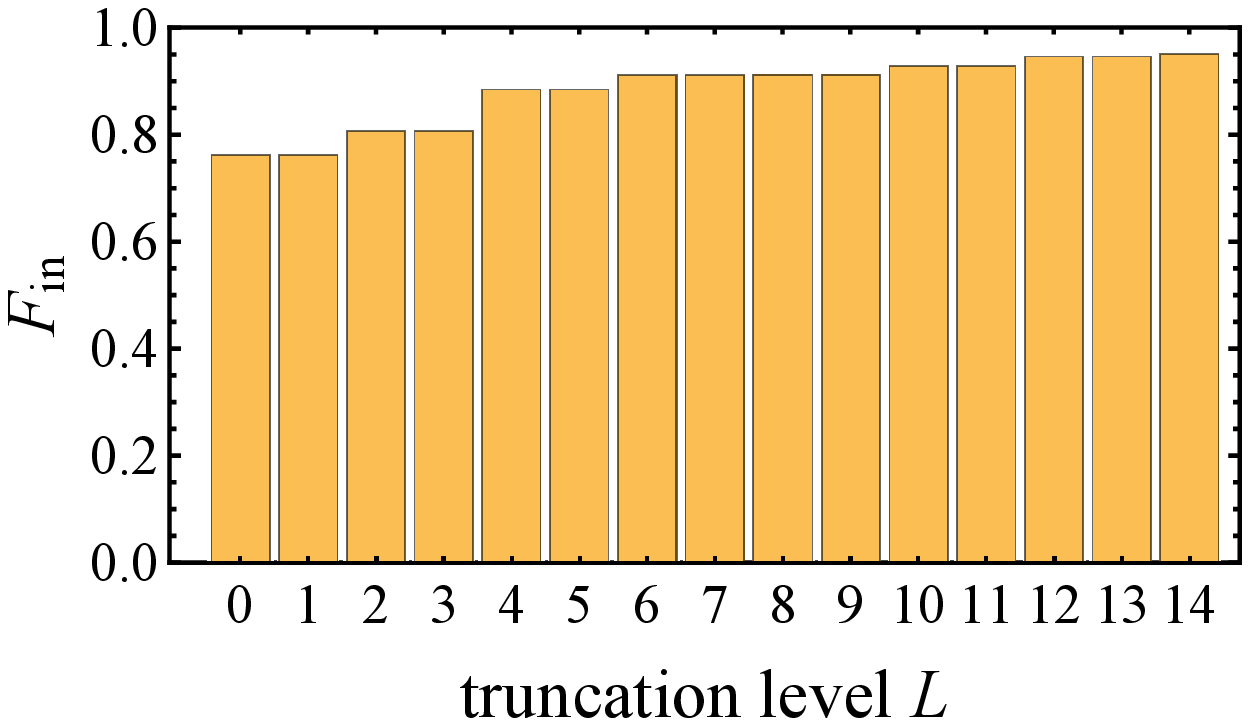}}
\caption{(Color online) (a) The OTOC in Eq. \eqref{OTOCn} for various pure initial Fock superposition states with driving strength $\xi=0.01$, and $x_0=\sqrt{\hbar/2m\omega}$ is vacuum position uncertainty; (b) the OTOC in Eq.~\eqref{OTOCn} for different Lyapunov exponent parameters (driving strengths) $\xi$; the initial pure state is $|\Psi_\text{in}\rangle=(|0\rangle+|1\rangle)/\sqrt{2}$; (c) the transmission and reflection probability 
in Eq.~\eqref{TR-pro} as function of ion displacement $x$, with fixed width $\Delta=\hbar\omega$, $\xi=0.01$, initial energy 
$\varepsilon_0=-0.5\hbar\omega$, and at time $t=2.86\,\mathrm{\mu\,s}$; (d)  
transmission and reflection probability as a function of time $t$, with 
ion displacement $x=1.35\,\mathrm{nm}$. 
(e) the fidelity of the prepared initial state \eqref{incident-wave} for various truncation 
levels $L$ of the Fock basis.
Everywhere, the oscillator frequency is taken to be 
$\omega_0/2\pi=10\,\mathrm{MHz}$.} \label{fig2}
\end{figure}

For the proposed experiment, we assume that the perturbation operator $W$ is an ion displacement measurement, 
$\hat{x}=\sqrt{\frac{\hbar}{2m\omega}}(\hat{a}+\hat{a}^\dagger)$, 
and
the initial pure state is prepared as, $| \Psi_\text{in}\rangle=\frac{1}{\sqrt{2}}(|n\rangle+|n+1\rangle)$, \cite{PhysRevLett.90.037902, PhysRevLett.92.113004}.
We then readily find that the OTOC is given by 
\begin{eqnarray}\label{OTOCn}
\nonumber
C(t)&=&\left|\langle \Psi_\text{in}|\,U^\dagger(t, 0)\hat{x}U(t, 0)|\Psi_\text{in}\rangle\right|^2
\\&=&
\frac{(n+1)\hbar}{2m\omega}\cosh^2(\lambda_Lt),
\end{eqnarray}
where $U(t, 0)=\exp[-iHt/\hbar]$. Measuring  OTOC then reduces to determining  directly 
accessible ion-motional quadratures  
\cite{ZMD, PhysRevLett.104.100503, PhysRevA.100.062111}. 
We find that
the OTOC exhibits exponential growth, $\propto e^{2\lambda_Lt}+(\text{terms growing more slowly than}\,e^{2\lambda_Lt})$, with a Lyapunov exponent 
$\lambda_L=\sqrt{\alpha/m}$. 
Fig.~\ref{fig2} shows the OTOC for various pure initial states and Lyapunov exponents.  

To observe the Hawking radiation associated to the S matrix \eqref{Smatrix}, 
we need to prepare a suitable incident wave packet. 
Specifically, we prepare the initial state of the ion's motion as a superposition of energy states, 
$|\Psi_\text{in}\rangle=\int\,d\varepsilon\,f(\varepsilon)|\varepsilon\rangle$.
When the distribution $f(\varepsilon)$ is assumed to be a Gaussian centered at an energy $\varepsilon_0$ with small width $\Delta$, i.e., 
$f(\varepsilon)=\frac{1}{\sqrt{2\pi^{3/2}\Delta}}\exp\big[-\frac{(\varepsilon-\varepsilon_0)^2}{2\Delta^2}\big]$, the incident wave packet can be written as \cite{BARTON1986322}
\begin{eqnarray}\label{incident-wave}
\Psi_\text{in}(x,t)=i{|x|^{-1/2}}e^{-i\varepsilon_0t-i\Phi_0}
F\big(t+\log[\sqrt{2}|x|]\big),
\end{eqnarray}
where $F(z)=(\Delta/\pi^{1/2})^{1/2}\exp\{-z^2\Delta^2/2\}$, $\Phi_0=\varepsilon_0\log\sqrt{2}|x|+x^2/2+\varphi_0/2+\pi/4$, 
with $\varphi_0=\arg\Gamma(\frac{1}{2}-i\varepsilon_0)$. 
Generating arbitrary harmonic-oscillator states has been experimentally demonstrated recently \cite{PhysRevLett.90.037902}. 
To prepare this wave packet in experiment, it is convenient 
to represent it in a phononic Fock basis, 
$|\Psi_\text{in}\rangle=\sum^\infty_{n=0}\langle\,n|\Psi_\text{in}\rangle|n\rangle$. 
We note that the overlap $\langle\,n|\Psi_\text{in}\rangle$
is nonzero only when $n$ is even. Using the experimental techniques of Refs.~\cite{PhysRevLett.90.037902, PhysRevLett.92.113004},  
the incident wave packet \eqref{incident-wave} can in principle  
be prepared experimentally. 
In Fig.~\ref{fig2}, we plot the fidelity 
$F_\text{in}=\sum^L_{n=0}|\langle\,n|\Psi_\text{in}\rangle|^2$ for preparing the incident wave packet, with $L$ being the truncation 
level in Fock space. 
For the initial state \eqref{incident-wave}, we derive an approximate analytical  expression 
for the transmitted and reflected probability densities as follows \cite{BARTON1986322}:
\begin{eqnarray}\label{TR-pro}
\nonumber
|\Psi_R|^2&=&|{\mathcal R}(\varepsilon_0)|^2\frac{1}{|x|}\left|F\left(t-\log[\sqrt{2}|x|]-\varphi^\prime(\varepsilon_0)\right)\right|^2,
\\ 
|\Psi_T|^2&=&|{\mathcal T}(\varepsilon_0)|^2\frac{1}{x}\left|F\left(t-\log[\sqrt{2}x]-\varphi^\prime(\varepsilon_0)\right)\right|^2. 
\end{eqnarray}
Here, $\varphi^\prime =d\varphi/d\varepsilon$, with  
$\varphi(\varepsilon)=\arg\Gamma(\frac{1}{2}-i\varepsilon)$.  
We have taken the units for mass, length, time, and energy as $2m$, 
$(\hbar/2m\omega)^{1/2}$, $\omega^{-1}$, and $\hbar\omega$, respectively. 
Integrating Eq.~\eqref{TR-pro} over space, we obtain the energy-dependent transmission and reflection probabilities in Eq.~\eqref{TR}.

In an experiment, the quantum mechanical ion motion can be detected by observing the evolution of its internal level populations  
according to the Hamiltonian 
$H_\mathrm{I}=\hbar\Omega\sigma_y(\hat{a}e^{i\phi/2}+\hat{a}^\dagger\,e^{-i\phi/2})$  \cite{PhysRevLett.76.1796}. 
This interaction can be implemented, in the Lamb-Dicke limit, 
by driving the ion with both blue and red sidebands, $H_b=\hbar\eta\Omega_b(\hat{a}^\dagger\sigma^+e^{i\phi_b}+\hat{a}\sigma^-e^{-i\phi_b})$ and 
$H_r=\hbar\eta\Omega_r(\hat{a}\sigma^+e^{i\phi_r}+\hat{a}^\dagger\sigma^-e^{-i\phi_r})$, respectively, 
and thus with the total Hamiltonian  $H_b+H_r
=\hbar\Omega(\hat{a}\,e^{i\phi_-}+\hat{a}^\dagger\,e^{-i\phi_-})(\sigma^+e^{i\phi_+}+\sigma^-e^{-i\phi_+})$. Here,  
we set $\eta\Omega_b=\eta\Omega_r=\Omega$, $2\phi_\pm=\phi_r\pm\phi_b$ by tuning the amplitude and phase of the applied driving field for the sidebands, and $\sigma^\pm=(\sigma_x\pm\,i\sigma_y)/2$.
To measure the scattering state, we first need to prepare the internal levels of the ion as  
$|+\rangle=1/\sqrt{2}(|\uparrow\rangle+|\downarrow\rangle)$,
and then apply the Raman-induced interaction coupling $H_\mathrm{I}$. 
We then record the probability $P_\downarrow(t)$ of occupation in $|\downarrow\rangle$.
The expected signal is $P_\downarrow(t)=\int\big(1-\sin(2\Omega\,tx)\big)|\Psi(x)|^2dx$ and 
the scattering probability $|\Psi(x)|^2$ in Eq. \eqref{TR-pro}
is then obtained by applying an inverse Fourier transform to $P_\downarrow(t)$.

Taking the initial energy, $\varepsilon_0<0$, 
we plot transmission and reflection probabilities in Fig.~\ref{fig2}. 
Negative $\epsilon_0$ implies that the particle
classically does not penetrate the barrier when coming from the left; 
Fig.~\ref{fig2} however clearly shows that both nonzero transmission and reflection probability 
exist simultaneously. 
The transmission probability, after integrating over position, 
satisfies a thermal distribution with temperature $T=\hbar\lambda_L/2\pi$. 
This mathematical relation between the temperature $T$ and the Lyapunov exponent $\lambda_L$ suggests that the conjectured MSS bound \cite{bound} is indeed satisfied for the ion's motion. 
We have thus shown that the ion trap IHO is the dual of a black hole. 

\begin{figure}[b]
\centering
\includegraphics[width=0.33\textwidth]{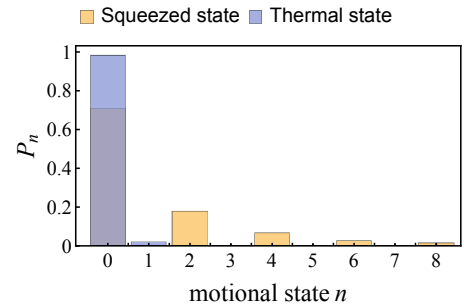}
\caption{(Color online) Phonon number distribution $P_{n}$ for a squeezed state (yellow)
and thermal state (light blue) at very low temperatures (average phonon number $\bar{n}=0.02$). 
The squeezing parameter is assumed to be $\lambda_Lt\approx0.88$. 
The vacuum squeezed state is created by the action of the Hamiltonian \eqref{Hamiltonian} on
the phonon vacuum. 
Although there are no phonons at the beginning, phonons are created during the evolution, as a result of analogue Hawking 
radiation.\label{fig3} }
\end{figure}

Rewriting Eq.~\eqref{Hamiltonian} with phononic annihilation and creation operators, 
we find $H=-\frac{1}{2}\hbar\lambda_L(\hat{a}^{\dagger2}+\hat{a}^2)$, which  
corresponds to a squeezing operation \cite{Loock}. 
We can thus alternatively interpret the analogue Hawking radiation via a squeezed state of the ion's motion. 
Phonon number increases as  
$\langle\hat{n}\rangle=\sinh^2(\lambda_Lt)$, and the squeezed vacuum state population distribution 
is restricted to even states, $P_{2n}=(2n)!\tanh^{2n}(\lambda_Lt)/(2^nn!)^2\cosh(\lambda_Lt)$  \cite{WANG20071}. 
The squeezed state can be detected by the evolution of the ion's internal levels under a Jaynes-Cummings interaction \cite{PhysRevLett.76.1796}. 
The population distribution is shown in Fig. \ref{fig3}, and 
compared with a slightly thermal state close to the vacuum.  
The increased population of higher Fock states   
are evidence of squeezing; only even states populated corresponds to the creation of pairs of phonons.

In conclusion, we propose using well established quantum optical tools 
for trapped ions to implement the IHO as the dual of a black hole. 
Chaotic quantum motion of the ion can be accessed by measuring the OTOC, 
which exhibits to leading order exponential growth, $\propto e^{2\lambda_Lt}$,  
with Lyapunov exponent $\lambda_L=\sqrt{\alpha/m}$, 
and analogue Hawking radiation temperature $T=\hbar\lambda_L/2\pi$. 

Numerous theoretical  approaches employed 
the IHO concept to establish links between quantum mechanics and general relativity see, e.g.,  
\cite{PhysRevLett.122.101603, PhysRevLett.123.156802, Betzios, Maldacena_2005, Morita:2018sen, PhysRevD.101.026021}.
Our IHO implementation proposal paves the way to experimentally 
explore these theoretical concepts in a highly controllable quantum optical environment.  


\begin{acknowledgments}
This work was supported by the National Key R\&D Program of China (Grant No. 2018YFA0306600), the CAS (Grants No. GJJSTD20170001 and No. QYZDY-SSW-SLH004), and Anhui Initiative in Quantum Information Technologies (Grant No. AHY050000).  
ZT was supported by the National Natural Science Foundation of 
China under Grant No. 11905218, and the CAS Key Laboratory for Research in Galaxies and Cosmology, Chinese Academy of Science (No. 18010203). YL. acknowledges support from the National Natural Science Foundation of China (Grant No. 11974330) and the Fundamental Research Funds for the Central Universities. 
URF has been supported by the NRF of Korea under Grant No.~2020R1A2C2008103. 

\end{acknowledgments}


\bibliography{ahr12}

\begin{thebibliography}{73}%
\makeatletter
\providecommand \@ifxundefined [1]{%
 \@ifx{#1\undefined}
}%
\providecommand \@ifnum [1]{%
 \ifnum #1\expandafter \@firstoftwo
 \else \expandafter \@secondoftwo
 \fi
}%
\providecommand \@ifx [1]{%
 \ifx #1\expandafter \@firstoftwo
 \else \expandafter \@secondoftwo
 \fi
}%
\providecommand \natexlab [1]{#1}%
\providecommand \enquote  [1]{``#1''}%
\providecommand \bibnamefont  [1]{#1}%
\providecommand \bibfnamefont [1]{#1}%
\providecommand \citenamefont [1]{#1}%
\providecommand \href@noop [0]{\@secondoftwo}%
\providecommand \href [0]{\begingroup \@sanitize@url \@href}%
\providecommand \@href[1]{\@@startlink{#1}\@@href}%
\providecommand \@@href[1]{\endgroup#1\@@endlink}%
\providecommand \@sanitize@url [0]{\catcode `\\12\catcode `\$12\catcode
  `\&12\catcode `\#12\catcode `\^12\catcode `\_12\catcode `\%12\relax}%
\providecommand \@@startlink[1]{}%
\providecommand \@@endlink[0]{}%
\providecommand \url  [0]{\begingroup\@sanitize@url \@url }%
\providecommand \@url [1]{\endgroup\@href {#1}{\urlprefix }}%
\providecommand \urlprefix  [0]{URL }%
\providecommand \Eprint [0]{\href }%
\providecommand \doibase [0]{http://dx.doi.org/}%
\providecommand \selectlanguage [0]{\@gobble}%
\providecommand \bibinfo  [0]{\@secondoftwo}%
\providecommand \bibfield  [0]{\@secondoftwo}%
\providecommand \translation [1]{[#1]}%
\providecommand \BibitemOpen [0]{}%
\providecommand \bibitemStop [0]{}%
\providecommand \bibitemNoStop [0]{.\EOS\space}%
\providecommand \EOS [0]{\spacefactor3000\relax}%
\providecommand \BibitemShut  [1]{\csname bibitem#1\endcsname}%
\let\auto@bib@innerbib\@empty
\bibitem [{\citenamefont {Hawking}(1974)}]{Hawking1}%
  \BibitemOpen
  \bibfield  {author} {\bibinfo {author} {\bibfnamefont {S.~W.}\ \bibnamefont
  {Hawking}},\ }\bibfield  {title} {\enquote {\bibinfo {title} {Black hole
  explosions?}}\ }\href {\doibase 10.1038/248030a0} {\bibfield  {journal}
  {\bibinfo  {journal} {Nature}\ }\textbf {\bibinfo {volume} {248}},\ \bibinfo
  {pages} {30--31} (\bibinfo {year} {1974})}\BibitemShut {NoStop}%
\bibitem [{\citenamefont {Hawking}(1975)}]{Hawking1975}%
  \BibitemOpen
  \bibfield  {author} {\bibinfo {author} {\bibfnamefont {S.~W.}\ \bibnamefont
  {Hawking}},\ }\bibfield  {title} {\enquote {\bibinfo {title} {Particle
  creation by black holes},}\ }\href {\doibase 10.1007/BF02345020} {\bibfield
  {journal} {\bibinfo  {journal} {Communications in Mathematical Physics}\
  }\textbf {\bibinfo {volume} {43}},\ \bibinfo {pages} {199--220} (\bibinfo
  {year} {1975})}\BibitemShut {NoStop}%
\bibitem [{\citenamefont {Unruh}(1981)}]{PhysRevLett.46.1351}%
  \BibitemOpen
  \bibfield  {author} {\bibinfo {author} {\bibfnamefont {W.~G.}\ \bibnamefont
  {Unruh}},\ }\bibfield  {title} {\enquote {\bibinfo {title} {{Experimental
  Black-Hole Evaporation?}}}\ }\href {\doibase 10.1103/PhysRevLett.46.1351}
  {\bibfield  {journal} {\bibinfo  {journal} {Phys. Rev. Lett.}\ }\textbf
  {\bibinfo {volume} {46}},\ \bibinfo {pages} {1351--1353} (\bibinfo {year}
  {1981})}\BibitemShut {NoStop}%
\bibitem [{\citenamefont {Barcel{\'o}}\ \emph {et~al.}(2011)\citenamefont
  {Barcel{\'o}}, \citenamefont {Liberati},\ and\ \citenamefont
  {Visser}}]{analogue-gravity}%
  \BibitemOpen
  \bibfield  {author} {\bibinfo {author} {\bibfnamefont {Carlos}\ \bibnamefont
  {Barcel{\'o}}}, \bibinfo {author} {\bibfnamefont {Stefano}\ \bibnamefont
  {Liberati}}, \ and\ \bibinfo {author} {\bibfnamefont {Matt}\ \bibnamefont
  {Visser}},\ }\bibfield  {title} {\enquote {\bibinfo {title} {{Analogue
  Gravity}},}\ }\href {\doibase 10.12942/lrr-2011-3} {\bibfield  {journal}
  {\bibinfo  {journal} {Living Reviews in Relativity}\ }\textbf {\bibinfo
  {volume} {14}},\ \bibinfo {pages} {3} (\bibinfo {year} {2011})}\BibitemShut
  {NoStop}%
\bibitem [{\citenamefont {Garay}\ \emph {et~al.}(2000)\citenamefont {Garay},
  \citenamefont {Anglin}, \citenamefont {Cirac},\ and\ \citenamefont
  {Zoller}}]{PhysRevLett.85.4643}%
  \BibitemOpen
  \bibfield  {author} {\bibinfo {author} {\bibfnamefont {L.~J.}\ \bibnamefont
  {Garay}}, \bibinfo {author} {\bibfnamefont {J.~R.}\ \bibnamefont {Anglin}},
  \bibinfo {author} {\bibfnamefont {J.~I.}\ \bibnamefont {Cirac}}, \ and\
  \bibinfo {author} {\bibfnamefont {P.}~\bibnamefont {Zoller}},\ }\bibfield
  {title} {\enquote {\bibinfo {title} {{Sonic Analog of Gravitational Black
  Holes in Bose-Einstein Condensates}},}\ }\href {\doibase
  10.1103/PhysRevLett.85.4643} {\bibfield  {journal} {\bibinfo  {journal}
  {Phys. Rev. Lett.}\ }\textbf {\bibinfo {volume} {85}},\ \bibinfo {pages}
  {4643--4647} (\bibinfo {year} {2000})}\BibitemShut {NoStop}%
\bibitem [{\citenamefont {Barcel\'o}\ \emph {et~al.}(2003)\citenamefont
  {Barcel\'o}, \citenamefont {Liberati},\ and\ \citenamefont {Visser}}]{BLV}%
  \BibitemOpen
  \bibfield  {author} {\bibinfo {author} {\bibfnamefont {Carlos}\ \bibnamefont
  {Barcel\'o}}, \bibinfo {author} {\bibfnamefont {S.}~\bibnamefont {Liberati}},
  \ and\ \bibinfo {author} {\bibfnamefont {Matt}\ \bibnamefont {Visser}},\
  }\bibfield  {title} {\enquote {\bibinfo {title} {{Probing semiclassical
  analog gravity in Bose-Einstein condensates with widely tunable
  interactions}},}\ }\href {\doibase 10.1103/PhysRevA.68.053613} {\bibfield
  {journal} {\bibinfo  {journal} {Phys. Rev. A}\ }\textbf {\bibinfo {volume}
  {68}},\ \bibinfo {pages} {053613} (\bibinfo {year} {2003})}\BibitemShut
  {NoStop}%
\bibitem [{\citenamefont {Fedichev}\ and\ \citenamefont
  {Fischer}(2003)}]{PhysRevLett.91.240407}%
  \BibitemOpen
  \bibfield  {author} {\bibinfo {author} {\bibfnamefont {Petr~O.}\ \bibnamefont
  {Fedichev}}\ and\ \bibinfo {author} {\bibfnamefont {Uwe~R.}\ \bibnamefont
  {Fischer}},\ }\bibfield  {title} {\enquote {\bibinfo {title}
  {{Gibbons-Hawking Effect in the Sonic de Sitter Space-Time of an Expanding
  Bose-Einstein-Condensed Gas}},}\ }\href {\doibase
  10.1103/PhysRevLett.91.240407} {\bibfield  {journal} {\bibinfo  {journal}
  {Phys. Rev. Lett.}\ }\textbf {\bibinfo {volume} {91}},\ \bibinfo {pages}
  {240407} (\bibinfo {year} {2003})}\BibitemShut {NoStop}%
\bibitem [{\citenamefont {Fedichev}\ and\ \citenamefont
  {Fischer}(2004)}]{PhysRevA.69.033602}%
  \BibitemOpen
  \bibfield  {author} {\bibinfo {author} {\bibfnamefont {Petr~O.}\ \bibnamefont
  {Fedichev}}\ and\ \bibinfo {author} {\bibfnamefont {Uwe~R.}\ \bibnamefont
  {Fischer}},\ }\bibfield  {title} {\enquote {\bibinfo {title}
  {{``Cosmological'' quasiparticle production in harmonically trapped
  superfluid gases}},}\ }\href {\doibase 10.1103/PhysRevA.69.033602} {\bibfield
   {journal} {\bibinfo  {journal} {Phys. Rev. A}\ }\textbf {\bibinfo {volume}
  {69}},\ \bibinfo {pages} {033602} (\bibinfo {year} {2004})}\BibitemShut
  {NoStop}%
\bibitem [{\citenamefont {Fischer}\ and\ \citenamefont
  {Sch\"utzhold}(2004)}]{Ralf}%
  \BibitemOpen
  \bibfield  {author} {\bibinfo {author} {\bibfnamefont {Uwe~R.}\ \bibnamefont
  {Fischer}}\ and\ \bibinfo {author} {\bibfnamefont {Ralf}\ \bibnamefont
  {Sch\"utzhold}},\ }\bibfield  {title} {\enquote {\bibinfo {title} {{Quantum
  simulation of cosmic inflation in two-component Bose-Einstein
  condensates}},}\ }\href {\doibase 10.1103/PhysRevA.70.063615} {\bibfield
  {journal} {\bibinfo  {journal} {Phys. Rev. A}\ }\textbf {\bibinfo {volume}
  {70}},\ \bibinfo {pages} {063615} (\bibinfo {year} {2004})}\BibitemShut
  {NoStop}%
\bibitem [{\citenamefont {Boiron}\ \emph {et~al.}(2015)\citenamefont {Boiron},
  \citenamefont {Fabbri}, \citenamefont {Larr\'e}, \citenamefont {Pavloff},
  \citenamefont {Westbrook},\ and\ \citenamefont {Zi\ifmmode~\acute{n}\else
  \'{n}\fi{}}}]{PhysRevLett.115.025301}%
  \BibitemOpen
  \bibfield  {author} {\bibinfo {author} {\bibfnamefont {D.}~\bibnamefont
  {Boiron}}, \bibinfo {author} {\bibfnamefont {A.}~\bibnamefont {Fabbri}},
  \bibinfo {author} {\bibfnamefont {P.-\'E.}\ \bibnamefont {Larr\'e}}, \bibinfo
  {author} {\bibfnamefont {N.}~\bibnamefont {Pavloff}}, \bibinfo {author}
  {\bibfnamefont {C.~I.}\ \bibnamefont {Westbrook}}, \ and\ \bibinfo {author}
  {\bibfnamefont {P.}~\bibnamefont {Zi\ifmmode~\acute{n}\else \'{n}\fi{}}},\
  }\bibfield  {title} {\enquote {\bibinfo {title} {{Quantum Signature of Analog
  Hawking Radiation in Momentum Space}},}\ }\href {\doibase
  10.1103/PhysRevLett.115.025301} {\bibfield  {journal} {\bibinfo  {journal}
  {Phys. Rev. Lett.}\ }\textbf {\bibinfo {volume} {115}},\ \bibinfo {pages}
  {025301} (\bibinfo {year} {2015})}\BibitemShut {NoStop}%
\bibitem [{\citenamefont {Carusotto}\ \emph {et~al.}(2008)\citenamefont
  {Carusotto}, \citenamefont {Fagnocchi}, \citenamefont {Recati}, \citenamefont
  {Balbinot},\ and\ \citenamefont {Fabbri}}]{Carusotto_2008}%
  \BibitemOpen
  \bibfield  {author} {\bibinfo {author} {\bibfnamefont {Iacopo}\ \bibnamefont
  {Carusotto}}, \bibinfo {author} {\bibfnamefont {Serena}\ \bibnamefont
  {Fagnocchi}}, \bibinfo {author} {\bibfnamefont {Alessio}\ \bibnamefont
  {Recati}}, \bibinfo {author} {\bibfnamefont {Roberto}\ \bibnamefont
  {Balbinot}}, \ and\ \bibinfo {author} {\bibfnamefont {Alessandro}\
  \bibnamefont {Fabbri}},\ }\bibfield  {title} {\enquote {\bibinfo {title}
  {{Numerical observation of Hawking radiation from acoustic black holes in
  atomic Bose{\textendash}Einstein condensates}},}\ }\href {\doibase
  10.1088/1367-2630/10/10/103001} {\bibfield  {journal} {\bibinfo  {journal}
  {New Journal of Physics}\ }\textbf {\bibinfo {volume} {10}},\ \bibinfo
  {pages} {103001} (\bibinfo {year} {2008})}\BibitemShut {NoStop}%
\bibitem [{\citenamefont {Steinhauer}(2014)}]{BEC1}%
  \BibitemOpen
  \bibfield  {author} {\bibinfo {author} {\bibfnamefont {Jeff}\ \bibnamefont
  {Steinhauer}},\ }\bibfield  {title} {\enquote {\bibinfo {title} {{Observation
  of self-amplifying Hawking radiation in an analogue black-hole laser}},}\
  }\href {\doibase 10.1038/nphys3104} {\bibfield  {journal} {\bibinfo
  {journal} {Nature Physics}\ }\textbf {\bibinfo {volume} {10}},\ \bibinfo
  {pages} {864--869} (\bibinfo {year} {2014})}\BibitemShut {NoStop}%
\bibitem [{\citenamefont {Steinhauer}(2016)}]{BEC2}%
  \BibitemOpen
  \bibfield  {author} {\bibinfo {author} {\bibfnamefont {Jeff}\ \bibnamefont
  {Steinhauer}},\ }\bibfield  {title} {\enquote {\bibinfo {title} {{Observation
  of quantum Hawking radiation and its entanglement in an analogue black
  hole}},}\ }\href {\doibase 10.1038/nphys3863} {\bibfield  {journal} {\bibinfo
   {journal} {Nature Physics}\ }\textbf {\bibinfo {volume} {12}},\ \bibinfo
  {pages} {959--965} (\bibinfo {year} {2016})}\BibitemShut {NoStop}%
\bibitem [{\citenamefont {Mu{\~n}oz~de Nova}\ \emph {et~al.}(2019)\citenamefont
  {Mu{\~n}oz~de Nova}, \citenamefont {Golubkov}, \citenamefont {Kolobov},\ and\
  \citenamefont {Steinhauer}}]{BEC3}%
  \BibitemOpen
  \bibfield  {author} {\bibinfo {author} {\bibfnamefont {Juan~Ram{\'o}n}\
  \bibnamefont {Mu{\~n}oz~de Nova}}, \bibinfo {author} {\bibfnamefont
  {Katrine}\ \bibnamefont {Golubkov}}, \bibinfo {author} {\bibfnamefont
  {Victor~I.}\ \bibnamefont {Kolobov}}, \ and\ \bibinfo {author} {\bibfnamefont
  {Jeff}\ \bibnamefont {Steinhauer}},\ }\bibfield  {title} {\enquote {\bibinfo
  {title} {{Observation of thermal Hawking radiation and its temperature in an
  analogue black hole}},}\ }\href {\doibase 10.1038/s41586-019-1241-0}
  {\bibfield  {journal} {\bibinfo  {journal} {Nature}\ }\textbf {\bibinfo
  {volume} {569}},\ \bibinfo {pages} {688--691} (\bibinfo {year}
  {2019})}\BibitemShut {NoStop}%
\bibitem [{\citenamefont {Michel}\ \emph {et~al.}(2016)\citenamefont {Michel},
  \citenamefont {Coupechoux},\ and\ \citenamefont {Parentani}}]{Renaud}%
  \BibitemOpen
  \bibfield  {author} {\bibinfo {author} {\bibfnamefont {Florent}\ \bibnamefont
  {Michel}}, \bibinfo {author} {\bibfnamefont
  {Jean-Fran\ifmmode\mbox{\c{c}}\else\c{c}\fi{}ois}\ \bibnamefont
  {Coupechoux}}, \ and\ \bibinfo {author} {\bibfnamefont {Renaud}\ \bibnamefont
  {Parentani}},\ }\bibfield  {title} {\enquote {\bibinfo {title} {{Phonon
  spectrum and correlations in a transonic flow of an atomic Bose gas}},}\
  }\href {\doibase 10.1103/PhysRevD.94.084027} {\bibfield  {journal} {\bibinfo
  {journal} {Phys. Rev. D}\ }\textbf {\bibinfo {volume} {94}},\ \bibinfo
  {pages} {084027} (\bibinfo {year} {2016})}\BibitemShut {NoStop}%
\bibitem [{\citenamefont {Isoard}\ and\ \citenamefont
  {Pavloff}(2020)}]{PhysRevLett.124.060401}%
  \BibitemOpen
  \bibfield  {author} {\bibinfo {author} {\bibfnamefont {M.}~\bibnamefont
  {Isoard}}\ and\ \bibinfo {author} {\bibfnamefont {N.}~\bibnamefont
  {Pavloff}},\ }\bibfield  {title} {\enquote {\bibinfo {title} {{Departing from
  Thermality of Analogue Hawking Radiation in a Bose-Einstein Condensate}},}\
  }\href {\doibase 10.1103/PhysRevLett.124.060401} {\bibfield  {journal}
  {\bibinfo  {journal} {Phys. Rev. Lett.}\ }\textbf {\bibinfo {volume} {124}},\
  \bibinfo {pages} {060401} (\bibinfo {year} {2020})}\BibitemShut {NoStop}%
\bibitem [{\citenamefont {Ch\"a}\ and\ \citenamefont
  {Fischer}(2017)}]{PhysRevLett.118.130404}%
  \BibitemOpen
  \bibfield  {author} {\bibinfo {author} {\bibfnamefont {Seok-Yeong}\
  \bibnamefont {Ch\"a}}\ and\ \bibinfo {author} {\bibfnamefont {Uwe~R.}\
  \bibnamefont {Fischer}},\ }\bibfield  {title} {\enquote {\bibinfo {title}
  {{Probing the Scale Invariance of the Inflationary Power Spectrum in
  Expanding Quasi-Two-Dimensional Dipolar Condensates}},}\ }\href {\doibase
  10.1103/PhysRevLett.118.130404} {\bibfield  {journal} {\bibinfo  {journal}
  {Phys. Rev. Lett.}\ }\textbf {\bibinfo {volume} {118}},\ \bibinfo {pages}
  {130404} (\bibinfo {year} {2017})}\BibitemShut {NoStop}%
\bibitem [{\citenamefont {Volkoff}\ and\ \citenamefont
  {Fischer}(2016)}]{Tyler}%
  \BibitemOpen
  \bibfield  {author} {\bibinfo {author} {\bibfnamefont {T.~J.}\ \bibnamefont
  {Volkoff}}\ and\ \bibinfo {author} {\bibfnamefont {Uwe~R.}\ \bibnamefont
  {Fischer}},\ }\bibfield  {title} {\enquote {\bibinfo {title} {{Quantum
  sine-Gordon dynamics on analogue curved spacetime in a weakly imperfect
  scalar Bose gas}},}\ }\href {\doibase 10.1103/PhysRevD.94.024051} {\bibfield
  {journal} {\bibinfo  {journal} {Phys. Rev. D}\ }\textbf {\bibinfo {volume}
  {94}},\ \bibinfo {pages} {024051} (\bibinfo {year} {2016})}\BibitemShut
  {NoStop}%
\bibitem [{\citenamefont {Philbin}\ \emph {et~al.}(2008)\citenamefont
  {Philbin}, \citenamefont {Kuklewicz}, \citenamefont {Robertson},
  \citenamefont {Hill}, \citenamefont {K{\"o}nig},\ and\ \citenamefont
  {Leonhardt}}]{Philbin1367}%
  \BibitemOpen
  \bibfield  {author} {\bibinfo {author} {\bibfnamefont {Thomas~G.}\
  \bibnamefont {Philbin}}, \bibinfo {author} {\bibfnamefont {Chris}\
  \bibnamefont {Kuklewicz}}, \bibinfo {author} {\bibfnamefont {Scott}\
  \bibnamefont {Robertson}}, \bibinfo {author} {\bibfnamefont {Stephen}\
  \bibnamefont {Hill}}, \bibinfo {author} {\bibfnamefont {Friedrich}\
  \bibnamefont {K{\"o}nig}}, \ and\ \bibinfo {author} {\bibfnamefont {Ulf}\
  \bibnamefont {Leonhardt}},\ }\bibfield  {title} {\enquote {\bibinfo {title}
  {{Fiber-Optical Analog of the Event Horizon}},}\ }\href {\doibase
  10.1126/science.1153625} {\bibfield  {journal} {\bibinfo  {journal}
  {Science}\ }\textbf {\bibinfo {volume} {319}},\ \bibinfo {pages} {1367--1370}
  (\bibinfo {year} {2008})}\BibitemShut {NoStop}%
\bibitem [{\citenamefont {Belgiorno}\ \emph {et~al.}(2010)\citenamefont
  {Belgiorno}, \citenamefont {Cacciatori}, \citenamefont {Clerici},
  \citenamefont {Gorini}, \citenamefont {Ortenzi}, \citenamefont {Rizzi},
  \citenamefont {Rubino}, \citenamefont {Sala},\ and\ \citenamefont
  {Faccio}}]{PhysRevLett.105.203901}%
  \BibitemOpen
  \bibfield  {author} {\bibinfo {author} {\bibfnamefont {F.}~\bibnamefont
  {Belgiorno}}, \bibinfo {author} {\bibfnamefont {S.~L.}\ \bibnamefont
  {Cacciatori}}, \bibinfo {author} {\bibfnamefont {M.}~\bibnamefont {Clerici}},
  \bibinfo {author} {\bibfnamefont {V.}~\bibnamefont {Gorini}}, \bibinfo
  {author} {\bibfnamefont {G.}~\bibnamefont {Ortenzi}}, \bibinfo {author}
  {\bibfnamefont {L.}~\bibnamefont {Rizzi}}, \bibinfo {author} {\bibfnamefont
  {E.}~\bibnamefont {Rubino}}, \bibinfo {author} {\bibfnamefont {V.~G.}\
  \bibnamefont {Sala}}, \ and\ \bibinfo {author} {\bibfnamefont
  {D.}~\bibnamefont {Faccio}},\ }\bibfield  {title} {\enquote {\bibinfo {title}
  {{Hawking Radiation from Ultrashort Laser Pulse Filaments}},}\ }\href
  {\doibase 10.1103/PhysRevLett.105.203901} {\bibfield  {journal} {\bibinfo
  {journal} {Phys. Rev. Lett.}\ }\textbf {\bibinfo {volume} {105}},\ \bibinfo
  {pages} {203901} (\bibinfo {year} {2010})}\BibitemShut {NoStop}%
\bibitem [{\citenamefont {Drori}\ \emph {et~al.}(2019)\citenamefont {Drori},
  \citenamefont {Rosenberg}, \citenamefont {Bermudez}, \citenamefont
  {Silberberg},\ and\ \citenamefont {Leonhardt}}]{PhysRevLett.122.010404}%
  \BibitemOpen
  \bibfield  {author} {\bibinfo {author} {\bibfnamefont {Jonathan}\
  \bibnamefont {Drori}}, \bibinfo {author} {\bibfnamefont {Yuval}\ \bibnamefont
  {Rosenberg}}, \bibinfo {author} {\bibfnamefont {David}\ \bibnamefont
  {Bermudez}}, \bibinfo {author} {\bibfnamefont {Yaron}\ \bibnamefont
  {Silberberg}}, \ and\ \bibinfo {author} {\bibfnamefont {Ulf}\ \bibnamefont
  {Leonhardt}},\ }\bibfield  {title} {\enquote {\bibinfo {title} {{Observation
  of Stimulated Hawking Radiation in an Optical Analogue}},}\ }\href {\doibase
  10.1103/PhysRevLett.122.010404} {\bibfield  {journal} {\bibinfo  {journal}
  {Phys. Rev. Lett.}\ }\textbf {\bibinfo {volume} {122}},\ \bibinfo {pages}
  {010404} (\bibinfo {year} {2019})}\BibitemShut {NoStop}%
\bibitem [{\citenamefont {Rold\'an-Molina}\ \emph {et~al.}(2017)\citenamefont
  {Rold\'an-Molina}, \citenamefont {Nunez},\ and\ \citenamefont
  {Duine}}]{PhysRevLett.118.061301}%
  \BibitemOpen
  \bibfield  {author} {\bibinfo {author} {\bibfnamefont {A.}~\bibnamefont
  {Rold\'an-Molina}}, \bibinfo {author} {\bibfnamefont {Alvaro~S.}\
  \bibnamefont {Nunez}}, \ and\ \bibinfo {author} {\bibfnamefont {R.~A.}\
  \bibnamefont {Duine}},\ }\bibfield  {title} {\enquote {\bibinfo {title}
  {{Magnonic Black Holes}},}\ }\href {\doibase 10.1103/PhysRevLett.118.061301}
  {\bibfield  {journal} {\bibinfo  {journal} {Phys. Rev. Lett.}\ }\textbf
  {\bibinfo {volume} {118}},\ \bibinfo {pages} {061301} (\bibinfo {year}
  {2017})}\BibitemShut {NoStop}%
\bibitem [{\citenamefont {Doornenbal}\ \emph {et~al.}(2019)\citenamefont
  {Doornenbal}, \citenamefont {Rold\'an-Molina}, \citenamefont {Nunez},\ and\
  \citenamefont {Duine}}]{PhysRevLett.122.037203}%
  \BibitemOpen
  \bibfield  {author} {\bibinfo {author} {\bibfnamefont {R.~J.}\ \bibnamefont
  {Doornenbal}}, \bibinfo {author} {\bibfnamefont {A.}~\bibnamefont
  {Rold\'an-Molina}}, \bibinfo {author} {\bibfnamefont {A.~S.}\ \bibnamefont
  {Nunez}}, \ and\ \bibinfo {author} {\bibfnamefont {R.~A.}\ \bibnamefont
  {Duine}},\ }\bibfield  {title} {\enquote {\bibinfo {title} {{Spin-Wave
  Amplification and Lasing Driven by Inhomogeneous Spin-Transfer Torques}},}\
  }\href {\doibase 10.1103/PhysRevLett.122.037203} {\bibfield  {journal}
  {\bibinfo  {journal} {Phys. Rev. Lett.}\ }\textbf {\bibinfo {volume} {122}},\
  \bibinfo {pages} {037203} (\bibinfo {year} {2019})}\BibitemShut {NoStop}%
\bibitem [{\citenamefont {Nation}\ \emph {et~al.}(2009)\citenamefont {Nation},
  \citenamefont {Blencowe}, \citenamefont {Rimberg},\ and\ \citenamefont
  {Buks}}]{PhysRevLett.103.087004}%
  \BibitemOpen
  \bibfield  {author} {\bibinfo {author} {\bibfnamefont {P.~D.}\ \bibnamefont
  {Nation}}, \bibinfo {author} {\bibfnamefont {M.~P.}\ \bibnamefont
  {Blencowe}}, \bibinfo {author} {\bibfnamefont {A.~J.}\ \bibnamefont
  {Rimberg}}, \ and\ \bibinfo {author} {\bibfnamefont {E.}~\bibnamefont
  {Buks}},\ }\bibfield  {title} {\enquote {\bibinfo {title} {{Analogue Hawking
  Radiation in a dc-SQUID Array Transmission Line}},}\ }\href {\doibase
  10.1103/PhysRevLett.103.087004} {\bibfield  {journal} {\bibinfo  {journal}
  {Phys. Rev. Lett.}\ }\textbf {\bibinfo {volume} {103}},\ \bibinfo {pages}
  {087004} (\bibinfo {year} {2009})}\BibitemShut {NoStop}%
\bibitem [{\citenamefont {Sab\'{\i}n}(2016)}]{PhysRevD.94.081501}%
  \BibitemOpen
  \bibfield  {author} {\bibinfo {author} {\bibfnamefont {Carlos}\ \bibnamefont
  {Sab\'{\i}n}},\ }\bibfield  {title} {\enquote {\bibinfo {title} {{Quantum
  simulation of traversable wormhole spacetimes in a dc-SQUID array}},}\ }\href
  {\doibase 10.1103/PhysRevD.94.081501} {\bibfield  {journal} {\bibinfo
  {journal} {Phys. Rev. D}\ }\textbf {\bibinfo {volume} {94}},\ \bibinfo
  {pages} {081501} (\bibinfo {year} {2016})}\BibitemShut {NoStop}%
\bibitem [{\citenamefont {Tian}\ \emph {et~al.}(2017)\citenamefont {Tian},
  \citenamefont {Jing},\ and\ \citenamefont {Dragan}}]{PhysRevD.95.125003}%
  \BibitemOpen
  \bibfield  {author} {\bibinfo {author} {\bibfnamefont {Zehua}\ \bibnamefont
  {Tian}}, \bibinfo {author} {\bibfnamefont {Jiliang}\ \bibnamefont {Jing}}, \
  and\ \bibinfo {author} {\bibfnamefont {Andrzej}\ \bibnamefont {Dragan}},\
  }\bibfield  {title} {\enquote {\bibinfo {title} {Analog cosmological particle
  generation in a superconducting circuit},}\ }\href {\doibase
  10.1103/PhysRevD.95.125003} {\bibfield  {journal} {\bibinfo  {journal} {Phys.
  Rev. D}\ }\textbf {\bibinfo {volume} {95}},\ \bibinfo {pages} {125003}
  (\bibinfo {year} {2017})}\BibitemShut {NoStop}%
\bibitem [{\citenamefont {Lang}\ and\ \citenamefont
  {Sch\"utzhold}(2019)}]{PhysRevD.100.065003}%
  \BibitemOpen
  \bibfield  {author} {\bibinfo {author} {\bibfnamefont {Sascha}\ \bibnamefont
  {Lang}}\ and\ \bibinfo {author} {\bibfnamefont {Ralf}\ \bibnamefont
  {Sch\"utzhold}},\ }\bibfield  {title} {\enquote {\bibinfo {title} {Analog of
  cosmological particle creation in electromagnetic waveguides},}\ }\href
  {\doibase 10.1103/PhysRevD.100.065003} {\bibfield  {journal} {\bibinfo
  {journal} {Phys. Rev. D}\ }\textbf {\bibinfo {volume} {100}},\ \bibinfo
  {pages} {065003} (\bibinfo {year} {2019})}\BibitemShut {NoStop}%
\bibitem [{\citenamefont {Tian}\ and\ \citenamefont
  {Du}(2019)}]{superconducting-circuit}%
  \BibitemOpen
  \bibfield  {author} {\bibinfo {author} {\bibfnamefont {Zehua}\ \bibnamefont
  {Tian}}\ and\ \bibinfo {author} {\bibfnamefont {Jiangfeng}\ \bibnamefont
  {Du}},\ }\bibfield  {title} {\enquote {\bibinfo {title} {{Analogue Hawking
  radiation and quantum soliton evaporation in a superconducting circuit}},}\
  }\href {\doibase 10.1140/epjc/s10052-019-7514-9} {\bibfield  {journal}
  {\bibinfo  {journal} {The European Physical Journal C}\ }\textbf {\bibinfo
  {volume} {79}},\ \bibinfo {pages} {994} (\bibinfo {year} {2019})}\BibitemShut
  {NoStop}%
\bibitem [{\citenamefont {Maldacena}(1999)}]{GGD1}%
  \BibitemOpen
  \bibfield  {author} {\bibinfo {author} {\bibfnamefont {Juan}\ \bibnamefont
  {Maldacena}},\ }\bibfield  {title} {\enquote {\bibinfo {title} {{The Large-N
  Limit of Superconformal Field Theories and Supergravity}},}\ }\href {\doibase
  10.1023/A:1026654312961} {\bibfield  {journal} {\bibinfo  {journal}
  {International Journal of Theoretical Physics}\ }\textbf {\bibinfo {volume}
  {38}},\ \bibinfo {pages} {1113--1133} (\bibinfo {year} {1999})}\BibitemShut
  {NoStop}%
\bibitem [{\citenamefont {Itzhaki}\ \emph {et~al.}(1998)\citenamefont
  {Itzhaki}, \citenamefont {Maldacena}, \citenamefont {Sonnenschein},\ and\
  \citenamefont {Yankielowicz}}]{PhysRevD.58.046004}%
  \BibitemOpen
  \bibfield  {author} {\bibinfo {author} {\bibfnamefont {Nissan}\ \bibnamefont
  {Itzhaki}}, \bibinfo {author} {\bibfnamefont {Juan~M.}\ \bibnamefont
  {Maldacena}}, \bibinfo {author} {\bibfnamefont {Jacob}\ \bibnamefont
  {Sonnenschein}}, \ and\ \bibinfo {author} {\bibfnamefont {Shimon}\
  \bibnamefont {Yankielowicz}},\ }\bibfield  {title} {\enquote {\bibinfo
  {title} {{Supergravity and the large $N$ limit of theories with sixteen
  supercharges}},}\ }\href {\doibase 10.1103/PhysRevD.58.046004} {\bibfield
  {journal} {\bibinfo  {journal} {Phys. Rev. D}\ }\textbf {\bibinfo {volume}
  {58}},\ \bibinfo {pages} {046004} (\bibinfo {year} {1998})}\BibitemShut
  {NoStop}%
\bibitem [{\citenamefont {Maldacena}\ \emph {et~al.}(2016)\citenamefont
  {Maldacena}, \citenamefont {Shenker},\ and\ \citenamefont
  {Stanford}}]{bound}%
  \BibitemOpen
  \bibfield  {author} {\bibinfo {author} {\bibfnamefont {Juan}\ \bibnamefont
  {Maldacena}}, \bibinfo {author} {\bibfnamefont {Stephen~H.}\ \bibnamefont
  {Shenker}}, \ and\ \bibinfo {author} {\bibfnamefont {Douglas}\ \bibnamefont
  {Stanford}},\ }\bibfield  {title} {\enquote {\bibinfo {title} {A bound on
  chaos},}\ }\href {\doibase 10.1007/JHEP08(2016)106} {\bibfield  {journal}
  {\bibinfo  {journal} {Journal of High Energy Physics}\ }\textbf {\bibinfo
  {volume} {2016}},\ \bibinfo {pages} {106} (\bibinfo {year}
  {2016})}\BibitemShut {NoStop}%
\bibitem [{\citenamefont {Shenker}\ and\ \citenamefont
  {Stanford}(2014)}]{QBC1}%
  \BibitemOpen
  \bibfield  {author} {\bibinfo {author} {\bibfnamefont {Stephen~H.}\
  \bibnamefont {Shenker}}\ and\ \bibinfo {author} {\bibfnamefont {Douglas}\
  \bibnamefont {Stanford}},\ }\bibfield  {title} {\enquote {\bibinfo {title}
  {Black holes and the butterfly effect},}\ }\href {\doibase
  10.1007/JHEP03(2014)067} {\bibfield  {journal} {\bibinfo  {journal} {Journal
  of High Energy Physics}\ }\textbf {\bibinfo {volume} {2014}},\ \bibinfo
  {pages} {67} (\bibinfo {year} {2014})}\BibitemShut {NoStop}%
\bibitem [{\citenamefont {Sekino}\ and\ \citenamefont
  {Susskind}(2008)}]{Sekino_2008}%
  \BibitemOpen
  \bibfield  {author} {\bibinfo {author} {\bibfnamefont {Yasuhiro}\
  \bibnamefont {Sekino}}\ and\ \bibinfo {author} {\bibfnamefont
  {L.}~\bibnamefont {Susskind}},\ }\bibfield  {title} {\enquote {\bibinfo
  {title} {Fast scramblers},}\ }\href {\doibase 10.1088/1126-6708/2008/10/065}
  {\bibfield  {journal} {\bibinfo  {journal} {Journal of High Energy Physics}\
  }\textbf {\bibinfo {volume} {2008}},\ \bibinfo {pages} {065--065} (\bibinfo
  {year} {2008})}\BibitemShut {NoStop}%
\bibitem [{\citenamefont {Franz}\ and\ \citenamefont
  {Rozali}(2018)}]{Franz:2018cqi}%
  \BibitemOpen
  \bibfield  {author} {\bibinfo {author} {\bibfnamefont {M.}~\bibnamefont
  {Franz}}\ and\ \bibinfo {author} {\bibfnamefont {M.}~\bibnamefont {Rozali}},\
  }\bibfield  {title} {\enquote {\bibinfo {title} {{Mimicking black hole event
  horizons in atomic and solid-state systems}},}\ }\href {\doibase
  10.1038/s41578-018-0058-z} {\bibfield  {journal} {\bibinfo  {journal} {Nature
  Rev. Mater.}\ }\textbf {\bibinfo {volume} {3}},\ \bibinfo {pages} {491--501}
  (\bibinfo {year} {2018})}\BibitemShut {NoStop}%
\bibitem [{\citenamefont {Morita}(2019)}]{PhysRevLett.122.101603}%
  \BibitemOpen
  \bibfield  {author} {\bibinfo {author} {\bibfnamefont {Takeshi}\ \bibnamefont
  {Morita}},\ }\bibfield  {title} {\enquote {\bibinfo {title} {{Thermal
  Emission from Semiclassical Dynamical Systems}},}\ }\href {\doibase
  10.1103/PhysRevLett.122.101603} {\bibfield  {journal} {\bibinfo  {journal}
  {Phys. Rev. Lett.}\ }\textbf {\bibinfo {volume} {122}},\ \bibinfo {pages}
  {101603} (\bibinfo {year} {2019})}\BibitemShut {NoStop}%
\bibitem [{\citenamefont {Betzios}\ \emph {et~al.}(2016)\citenamefont
  {Betzios}, \citenamefont {Gaddam},\ and\ \citenamefont
  {Papadoulaki}}]{Betzios}%
  \BibitemOpen
  \bibfield  {author} {\bibinfo {author} {\bibfnamefont {Panagiotis}\
  \bibnamefont {Betzios}}, \bibinfo {author} {\bibfnamefont {Nava}\
  \bibnamefont {Gaddam}}, \ and\ \bibinfo {author} {\bibfnamefont {Olga}\
  \bibnamefont {Papadoulaki}},\ }\bibfield  {title} {\enquote {\bibinfo {title}
  {{The black hole S-Matrix from quantum mechanics}},}\ }\href {\doibase
  10.1007/JHEP11(2016)131} {\bibfield  {journal} {\bibinfo  {journal} {Journal
  of High Energy Physics}\ }\textbf {\bibinfo {volume} {2016}},\ \bibinfo
  {pages} {131} (\bibinfo {year} {2016})}\BibitemShut {NoStop}%
\bibitem [{\citenamefont {Maldacena}\ and\ \citenamefont
  {Seiberg}(2005)}]{Maldacena_2005}%
  \BibitemOpen
  \bibfield  {author} {\bibinfo {author} {\bibfnamefont {Juan}\ \bibnamefont
  {Maldacena}}\ and\ \bibinfo {author} {\bibfnamefont {Nathan}\ \bibnamefont
  {Seiberg}},\ }\bibfield  {title} {\enquote {\bibinfo {title} {Flux-vacua in
  two dimensional string theory},}\ }\href {\doibase
  10.1088/1126-6708/2005/09/077} {\bibfield  {journal} {\bibinfo  {journal}
  {Journal of High Energy Physics}\ }\textbf {\bibinfo {volume} {2005}},\
  \bibinfo {pages} {077--077} (\bibinfo {year} {2005})}\BibitemShut {NoStop}%
\bibitem [{\citenamefont {{Larkin}}\ and\ \citenamefont
  {{Ovchinnikov}}(1969)}]{1969JETP1200L}%
  \BibitemOpen
  \bibfield  {author} {\bibinfo {author} {\bibfnamefont {A.~I.}\ \bibnamefont
  {{Larkin}}}\ and\ \bibinfo {author} {\bibfnamefont {Yu.~N.}\ \bibnamefont
  {{Ovchinnikov}}},\ }\bibfield  {title} {\enquote {\bibinfo {title}
  {{{Quasiclassical Method in the Theory of Superconductivity}}},}\ }\href@noop
  {} {\bibfield  {journal} {\bibinfo  {journal} {Soviet Journal of Experimental
  and Theoretical Physics}\ }\textbf {\bibinfo {volume} {28}},\ \bibinfo
  {pages} {1200} (\bibinfo {year} {1969})}\BibitemShut {NoStop}%
\bibitem [{\citenamefont {Lamata}\ \emph {et~al.}(2007)\citenamefont {Lamata},
  \citenamefont {Le\'on}, \citenamefont {Sch\"atz},\ and\ \citenamefont
  {Solano}}]{PhysRevLett.98.253005}%
  \BibitemOpen
  \bibfield  {author} {\bibinfo {author} {\bibfnamefont {L.}~\bibnamefont
  {Lamata}}, \bibinfo {author} {\bibfnamefont {J.}~\bibnamefont {Le\'on}},
  \bibinfo {author} {\bibfnamefont {T.}~\bibnamefont {Sch\"atz}}, \ and\
  \bibinfo {author} {\bibfnamefont {E.}~\bibnamefont {Solano}},\ }\bibfield
  {title} {\enquote {\bibinfo {title} {{Dirac Equation and Quantum Relativistic
  Effects in a Single Trapped Ion}},}\ }\href {\doibase
  10.1103/PhysRevLett.98.253005} {\bibfield  {journal} {\bibinfo  {journal}
  {Phys. Rev. Lett.}\ }\textbf {\bibinfo {volume} {98}},\ \bibinfo {pages}
  {253005} (\bibinfo {year} {2007})}\BibitemShut {NoStop}%
\bibitem [{\citenamefont {Gerritsma}\ \emph {et~al.}(2010)\citenamefont
  {Gerritsma}, \citenamefont {Kirchmair}, \citenamefont {Z{\"a}hringer},
  \citenamefont {Solano}, \citenamefont {Blatt},\ and\ \citenamefont
  {Roos}}]{ZMD}%
  \BibitemOpen
  \bibfield  {author} {\bibinfo {author} {\bibfnamefont {R.}~\bibnamefont
  {Gerritsma}}, \bibinfo {author} {\bibfnamefont {G.}~\bibnamefont
  {Kirchmair}}, \bibinfo {author} {\bibfnamefont {F.}~\bibnamefont
  {Z{\"a}hringer}}, \bibinfo {author} {\bibfnamefont {E.}~\bibnamefont
  {Solano}}, \bibinfo {author} {\bibfnamefont {R.}~\bibnamefont {Blatt}}, \
  and\ \bibinfo {author} {\bibfnamefont {C.~F.}\ \bibnamefont {Roos}},\
  }\bibfield  {title} {\enquote {\bibinfo {title} {{Quantum simulation of the
  Dirac equation}},}\ }\href {\doibase 10.1038/nature08688} {\bibfield
  {journal} {\bibinfo  {journal} {Nature}\ }\textbf {\bibinfo {volume} {463}},\
  \bibinfo {pages} {68--71} (\bibinfo {year} {2010})}\BibitemShut {NoStop}%
\bibitem [{\citenamefont {Pedernales}\ \emph {et~al.}(2018)\citenamefont
  {Pedernales}, \citenamefont {Beau}, \citenamefont {Pittman}, \citenamefont
  {Egusquiza}, \citenamefont {Lamata}, \citenamefont {Solano},\ and\
  \citenamefont {del Campo}}]{PhysRevLett.120.160403}%
  \BibitemOpen
  \bibfield  {author} {\bibinfo {author} {\bibfnamefont {J.~S.}\ \bibnamefont
  {Pedernales}}, \bibinfo {author} {\bibfnamefont {M.}~\bibnamefont {Beau}},
  \bibinfo {author} {\bibfnamefont {S.~M.}\ \bibnamefont {Pittman}}, \bibinfo
  {author} {\bibfnamefont {I.~L.}\ \bibnamefont {Egusquiza}}, \bibinfo {author}
  {\bibfnamefont {L.}~\bibnamefont {Lamata}}, \bibinfo {author} {\bibfnamefont
  {E.}~\bibnamefont {Solano}}, \ and\ \bibinfo {author} {\bibfnamefont
  {A.}~\bibnamefont {del Campo}},\ }\bibfield  {title} {\enquote {\bibinfo
  {title} {{Dirac Equation in ($1+1$)-Dimensional Curved Spacetime and the
  Multiphoton Quantum Rabi Model}},}\ }\href {\doibase
  10.1103/PhysRevLett.120.160403} {\bibfield  {journal} {\bibinfo  {journal}
  {Phys. Rev. Lett.}\ }\textbf {\bibinfo {volume} {120}},\ \bibinfo {pages}
  {160403} (\bibinfo {year} {2018})}\BibitemShut {NoStop}%
\bibitem [{\citenamefont {Gerritsma}\ \emph {et~al.}(2011)\citenamefont
  {Gerritsma}, \citenamefont {Lanyon}, \citenamefont {Kirchmair}, \citenamefont
  {Z\"ahringer}, \citenamefont {Hempel}, \citenamefont {Casanova},
  \citenamefont {Garc\'{\i}a-Ripoll}, \citenamefont {Solano}, \citenamefont
  {Blatt},\ and\ \citenamefont {Roos}}]{PhysRevLett.106.060503}%
  \BibitemOpen
  \bibfield  {author} {\bibinfo {author} {\bibfnamefont {R.}~\bibnamefont
  {Gerritsma}}, \bibinfo {author} {\bibfnamefont {B.~P.}\ \bibnamefont
  {Lanyon}}, \bibinfo {author} {\bibfnamefont {G.}~\bibnamefont {Kirchmair}},
  \bibinfo {author} {\bibfnamefont {F.}~\bibnamefont {Z\"ahringer}}, \bibinfo
  {author} {\bibfnamefont {C.}~\bibnamefont {Hempel}}, \bibinfo {author}
  {\bibfnamefont {J.}~\bibnamefont {Casanova}}, \bibinfo {author}
  {\bibfnamefont {J.~J.}\ \bibnamefont {Garc\'{\i}a-Ripoll}}, \bibinfo {author}
  {\bibfnamefont {E.}~\bibnamefont {Solano}}, \bibinfo {author} {\bibfnamefont
  {R.}~\bibnamefont {Blatt}}, \ and\ \bibinfo {author} {\bibfnamefont {C.~F.}\
  \bibnamefont {Roos}},\ }\bibfield  {title} {\enquote {\bibinfo {title}
  {{Quantum Simulation of the Klein Paradox with Trapped Ions}},}\ }\href
  {\doibase 10.1103/PhysRevLett.106.060503} {\bibfield  {journal} {\bibinfo
  {journal} {Phys. Rev. Lett.}\ }\textbf {\bibinfo {volume} {106}},\ \bibinfo
  {pages} {060503} (\bibinfo {year} {2011})}\BibitemShut {NoStop}%
\bibitem [{\citenamefont {Casanova}\ \emph {et~al.}(2010)\citenamefont
  {Casanova}, \citenamefont {Garc\'{\i}a-Ripoll}, \citenamefont {Gerritsma},
  \citenamefont {Roos},\ and\ \citenamefont {Solano}}]{PhysRevA.82.020101}%
  \BibitemOpen
  \bibfield  {author} {\bibinfo {author} {\bibfnamefont {J.}~\bibnamefont
  {Casanova}}, \bibinfo {author} {\bibfnamefont {J.~J.}\ \bibnamefont
  {Garc\'{\i}a-Ripoll}}, \bibinfo {author} {\bibfnamefont {R.}~\bibnamefont
  {Gerritsma}}, \bibinfo {author} {\bibfnamefont {C.~F.}\ \bibnamefont {Roos}},
  \ and\ \bibinfo {author} {\bibfnamefont {E.}~\bibnamefont {Solano}},\
  }\bibfield  {title} {\enquote {\bibinfo {title} {{Klein tunneling and Dirac
  potentials in trapped ions}},}\ }\href {\doibase 10.1103/PhysRevA.82.020101}
  {\bibfield  {journal} {\bibinfo  {journal} {Phys. Rev. A}\ }\textbf {\bibinfo
  {volume} {82}},\ \bibinfo {pages} {020101} (\bibinfo {year}
  {2010})}\BibitemShut {NoStop}%
\bibitem [{\citenamefont {Alsing}\ \emph {et~al.}(2005)\citenamefont {Alsing},
  \citenamefont {Dowling},\ and\ \citenamefont
  {Milburn}}]{PhysRevLett.94.220401}%
  \BibitemOpen
  \bibfield  {author} {\bibinfo {author} {\bibfnamefont {Paul~M.}\ \bibnamefont
  {Alsing}}, \bibinfo {author} {\bibfnamefont {Jonathan~P.}\ \bibnamefont
  {Dowling}}, \ and\ \bibinfo {author} {\bibfnamefont {G.~J.}\ \bibnamefont
  {Milburn}},\ }\bibfield  {title} {\enquote {\bibinfo {title} {{Ion Trap
  Simulations of Quantum Fields in an Expanding Universe}},}\ }\href {\doibase
  10.1103/PhysRevLett.94.220401} {\bibfield  {journal} {\bibinfo  {journal}
  {Phys. Rev. Lett.}\ }\textbf {\bibinfo {volume} {94}},\ \bibinfo {pages}
  {220401} (\bibinfo {year} {2005})}\BibitemShut {NoStop}%
\bibitem [{\citenamefont {Sch\"utzhold}\ \emph {et~al.}(2007)\citenamefont
  {Sch\"utzhold}, \citenamefont {Uhlmann}, \citenamefont {Petersen},
  \citenamefont {Schmitz}, \citenamefont {Friedenauer},\ and\ \citenamefont
  {Sch\"atz}}]{PhysRevLett.99.201301}%
  \BibitemOpen
  \bibfield  {author} {\bibinfo {author} {\bibfnamefont {Ralf}\ \bibnamefont
  {Sch\"utzhold}}, \bibinfo {author} {\bibfnamefont {Michael}\ \bibnamefont
  {Uhlmann}}, \bibinfo {author} {\bibfnamefont {Lutz}\ \bibnamefont
  {Petersen}}, \bibinfo {author} {\bibfnamefont {Hector}\ \bibnamefont
  {Schmitz}}, \bibinfo {author} {\bibfnamefont {Axel}\ \bibnamefont
  {Friedenauer}}, \ and\ \bibinfo {author} {\bibfnamefont {Tobias}\
  \bibnamefont {Sch\"atz}},\ }\bibfield  {title} {\enquote {\bibinfo {title}
  {{Analogue of Cosmological Particle Creation in an Ion Trap}},}\ }\href
  {\doibase 10.1103/PhysRevLett.99.201301} {\bibfield  {journal} {\bibinfo
  {journal} {Phys. Rev. Lett.}\ }\textbf {\bibinfo {volume} {99}},\ \bibinfo
  {pages} {201301} (\bibinfo {year} {2007})}\BibitemShut {NoStop}%
\bibitem [{\citenamefont {Wittemer}\ \emph {et~al.}(2019)\citenamefont
  {Wittemer}, \citenamefont {Hakelberg}, \citenamefont {Kiefer}, \citenamefont
  {Schr\"oder}, \citenamefont {Fey}, \citenamefont {Sch\"utzhold},
  \citenamefont {Warring},\ and\ \citenamefont
  {Schaetz}}]{PhysRevLett.123.180502}%
  \BibitemOpen
  \bibfield  {author} {\bibinfo {author} {\bibfnamefont {Matthias}\
  \bibnamefont {Wittemer}}, \bibinfo {author} {\bibfnamefont {Frederick}\
  \bibnamefont {Hakelberg}}, \bibinfo {author} {\bibfnamefont {Philip}\
  \bibnamefont {Kiefer}}, \bibinfo {author} {\bibfnamefont {Jan-Philipp}\
  \bibnamefont {Schr\"oder}}, \bibinfo {author} {\bibfnamefont {Christian}\
  \bibnamefont {Fey}}, \bibinfo {author} {\bibfnamefont {Ralf}\ \bibnamefont
  {Sch\"utzhold}}, \bibinfo {author} {\bibfnamefont {Ulrich}\ \bibnamefont
  {Warring}}, \ and\ \bibinfo {author} {\bibfnamefont {Tobias}\ \bibnamefont
  {Schaetz}},\ }\bibfield  {title} {\enquote {\bibinfo {title} {{Phonon Pair
  Creation by Inflating Quantum Fluctuations in an Ion Trap}},}\ }\href
  {\doibase 10.1103/PhysRevLett.123.180502} {\bibfield  {journal} {\bibinfo
  {journal} {Phys. Rev. Lett.}\ }\textbf {\bibinfo {volume} {123}},\ \bibinfo
  {pages} {180502} (\bibinfo {year} {2019})}\BibitemShut {NoStop}%
\bibitem [{\citenamefont {Menicucci}\ \emph {et~al.}(2010)\citenamefont
  {Menicucci}, \citenamefont {Olson},\ and\ \citenamefont
  {Milburn}}]{Menicucci_2010}%
  \BibitemOpen
  \bibfield  {author} {\bibinfo {author} {\bibfnamefont {Nicolas~C.}\
  \bibnamefont {Menicucci}}, \bibinfo {author} {\bibfnamefont {S.~Jay}\
  \bibnamefont {Olson}}, \ and\ \bibinfo {author} {\bibfnamefont {Gerard~J.}\
  \bibnamefont {Milburn}},\ }\bibfield  {title} {\enquote {\bibinfo {title}
  {Simulating quantum effects of cosmological expansion using a static ion
  trap},}\ }\href {\doibase 10.1088/1367-2630/12/9/095019} {\bibfield
  {journal} {\bibinfo  {journal} {New Journal of Physics}\ }\textbf {\bibinfo
  {volume} {12}},\ \bibinfo {pages} {095019} (\bibinfo {year}
  {2010})}\BibitemShut {NoStop}%
\bibitem [{\citenamefont {Hashimoto}\ and\ \citenamefont
  {Tanahashi}(2017)}]{PhysRevD.95.024007}%
  \BibitemOpen
  \bibfield  {author} {\bibinfo {author} {\bibfnamefont {Koji}\ \bibnamefont
  {Hashimoto}}\ and\ \bibinfo {author} {\bibfnamefont {Norihiro}\ \bibnamefont
  {Tanahashi}},\ }\bibfield  {title} {\enquote {\bibinfo {title} {Universality
  in chaos of particle motion near black hole horizon},}\ }\href {\doibase
  10.1103/PhysRevD.95.024007} {\bibfield  {journal} {\bibinfo  {journal} {Phys.
  Rev. D}\ }\textbf {\bibinfo {volume} {95}},\ \bibinfo {pages} {024007}
  (\bibinfo {year} {2017})}\BibitemShut {NoStop}%
\bibitem [{\citenamefont {Gibbons}\ and\ \citenamefont
  {Hawking}(1977)}]{Gibbons}%
  \BibitemOpen
  \bibfield  {author} {\bibinfo {author} {\bibfnamefont {G.~W.}\ \bibnamefont
  {Gibbons}}\ and\ \bibinfo {author} {\bibfnamefont {S.~W.}\ \bibnamefont
  {Hawking}},\ }\bibfield  {title} {\enquote {\bibinfo {title} {Action
  integrals and partition functions in quantum gravity},}\ }\href {\doibase
  10.1103/PhysRevD.15.2752} {\bibfield  {journal} {\bibinfo  {journal} {Phys.
  Rev. D}\ }\textbf {\bibinfo {volume} {15}},\ \bibinfo {pages} {2752--2756}
  (\bibinfo {year} {1977})}\BibitemShut {NoStop}%
\bibitem [{\citenamefont {Parikh}\ and\ \citenamefont
  {Wilczek}(2000)}]{Parikh}%
  \BibitemOpen
  \bibfield  {author} {\bibinfo {author} {\bibfnamefont {Maulik~K.}\
  \bibnamefont {Parikh}}\ and\ \bibinfo {author} {\bibfnamefont {Frank}\
  \bibnamefont {Wilczek}},\ }\bibfield  {title} {\enquote {\bibinfo {title}
  {{Hawking Radiation As Tunneling}},}\ }\href {\doibase
  10.1103/PhysRevLett.85.5042} {\bibfield  {journal} {\bibinfo  {journal}
  {Phys. Rev. Lett.}\ }\textbf {\bibinfo {volume} {85}},\ \bibinfo {pages}
  {5042--5045} (\bibinfo {year} {2000})}\BibitemShut {NoStop}%
\bibitem [{\citenamefont {Robinson}\ and\ \citenamefont
  {Wilczek}(2005)}]{Robinson}%
  \BibitemOpen
  \bibfield  {author} {\bibinfo {author} {\bibfnamefont {Sean~P.}\ \bibnamefont
  {Robinson}}\ and\ \bibinfo {author} {\bibfnamefont {Frank}\ \bibnamefont
  {Wilczek}},\ }\bibfield  {title} {\enquote {\bibinfo {title} {{Relationship
  between Hawking Radiation and Gravitational Anomalies}},}\ }\href {\doibase
  10.1103/PhysRevLett.95.011303} {\bibfield  {journal} {\bibinfo  {journal}
  {Phys. Rev. Lett.}\ }\textbf {\bibinfo {volume} {95}},\ \bibinfo {pages}
  {011303} (\bibinfo {year} {2005})}\BibitemShut {NoStop}%
\bibitem [{\citenamefont {Iso}\ \emph {et~al.}(2006)\citenamefont {Iso},
  \citenamefont {Umetsu},\ and\ \citenamefont {Wilczek}}]{Satoshi}%
  \BibitemOpen
  \bibfield  {author} {\bibinfo {author} {\bibfnamefont {Satoshi}\ \bibnamefont
  {Iso}}, \bibinfo {author} {\bibfnamefont {Hiroshi}\ \bibnamefont {Umetsu}}, \
  and\ \bibinfo {author} {\bibfnamefont {Frank}\ \bibnamefont {Wilczek}},\
  }\bibfield  {title} {\enquote {\bibinfo {title} {{Hawking Radiation from
  Charged Black Holes via Gauge and Gravitational Anomalies}},}\ }\href
  {\doibase 10.1103/PhysRevLett.96.151302} {\bibfield  {journal} {\bibinfo
  {journal} {Phys. Rev. Lett.}\ }\textbf {\bibinfo {volume} {96}},\ \bibinfo
  {pages} {151302} (\bibinfo {year} {2006})}\BibitemShut {NoStop}%
\bibitem [{\citenamefont {Peet}(2001)}]{Peet:2000hn}%
  \BibitemOpen
  \bibfield  {author} {\bibinfo {author} {\bibfnamefont {Amanda~W.}\
  \bibnamefont {Peet}},\ }\enquote {\bibinfo {title} {{TASI Lectures on Black
  Holes in String Theory}},}\ in\ \href {\doibase 10.1142/9789812799630_0003}
  {\emph {\bibinfo {booktitle} {Strings, Branes and Gravity}}}\ (\bibinfo
  {publisher} {World Scientific},\ \bibinfo {year} {2001})\ pp.\ \bibinfo
  {pages} {353--433}\BibitemShut {NoStop}%
\bibitem [{\citenamefont {Strominger}\ and\ \citenamefont
  {Vafa}(1996)}]{STROMINGER199699}%
  \BibitemOpen
  \bibfield  {author} {\bibinfo {author} {\bibfnamefont {Andrew}\ \bibnamefont
  {Strominger}}\ and\ \bibinfo {author} {\bibfnamefont {Cumrun}\ \bibnamefont
  {Vafa}},\ }\bibfield  {title} {\enquote {\bibinfo {title} {{{Microscopic
  origin of the Bekenstein-Hawking entropy}}},}\ }\href {\doibase
  https://doi.org/10.1016/0370-2693(96)00345-0} {\bibfield  {journal} {\bibinfo
   {journal} {Physics Letters B}\ }\textbf {\bibinfo {volume} {379}},\ \bibinfo
  {pages} {99--104} (\bibinfo {year} {1996})}\BibitemShut {NoStop}%
\bibitem [{\citenamefont {Yu}\ and\ \citenamefont
  {Zhang}(2008)}]{PhysRevD.77.024031}%
  \BibitemOpen
  \bibfield  {author} {\bibinfo {author} {\bibfnamefont {Hongwei}\ \bibnamefont
  {Yu}}\ and\ \bibinfo {author} {\bibfnamefont {Jialin}\ \bibnamefont
  {Zhang}},\ }\bibfield  {title} {\enquote {\bibinfo {title} {{Understanding
  Hawking radiation in the framework of open quantum systems}},}\ }\href
  {\doibase 10.1103/PhysRevD.77.024031} {\bibfield  {journal} {\bibinfo
  {journal} {Phys. Rev. D}\ }\textbf {\bibinfo {volume} {77}},\ \bibinfo
  {pages} {024031} (\bibinfo {year} {2008})}\BibitemShut {NoStop}%
\bibitem [{\citenamefont {Kerner}\ and\ \citenamefont
  {Mann}(2008)}]{Kerner2008}%
  \BibitemOpen
  \bibfield  {author} {\bibinfo {author} {\bibfnamefont {Ryan}\ \bibnamefont
  {Kerner}}\ and\ \bibinfo {author} {\bibfnamefont {R.~B.}\ \bibnamefont
  {Mann}},\ }\bibfield  {title} {\enquote {\bibinfo {title} {Fermions
  tunnelling from black holes},}\ }\href {\doibase
  10.1088/0264-9381/25/9/095014} {\bibfield  {journal} {\bibinfo  {journal}
  {Classical and Quantum Gravity}\ }\textbf {\bibinfo {volume} {25}},\ \bibinfo
  {pages} {095014} (\bibinfo {year} {2008})}\BibitemShut {NoStop}%
\bibitem [{\citenamefont {Leibfried}\ \emph {et~al.}(2003)\citenamefont
  {Leibfried}, \citenamefont {Blatt}, \citenamefont {Monroe},\ and\
  \citenamefont {Wineland}}]{RevModPhys.75.281}%
  \BibitemOpen
  \bibfield  {author} {\bibinfo {author} {\bibfnamefont {D.}~\bibnamefont
  {Leibfried}}, \bibinfo {author} {\bibfnamefont {R.}~\bibnamefont {Blatt}},
  \bibinfo {author} {\bibfnamefont {C.}~\bibnamefont {Monroe}}, \ and\ \bibinfo
  {author} {\bibfnamefont {D.}~\bibnamefont {Wineland}},\ }\bibfield  {title}
  {\enquote {\bibinfo {title} {Quantum dynamics of single trapped ions},}\
  }\href {\doibase 10.1103/RevModPhys.75.281} {\bibfield  {journal} {\bibinfo
  {journal} {Rev. Mod. Phys.}\ }\textbf {\bibinfo {volume} {75}},\ \bibinfo
  {pages} {281--324} (\bibinfo {year} {2003})}\BibitemShut {NoStop}%
\bibitem [{\citenamefont {Blocher}\ \emph {et~al.}(2020)\citenamefont
  {Blocher}, \citenamefont {Asaad}, \citenamefont {Mourik}, \citenamefont
  {Johnson}, \citenamefont {Morello},\ and\ \citenamefont
  {M\o{}lmer}}]{blocher2020measuring}%
  \BibitemOpen
  \bibfield  {author} {\bibinfo {author} {\bibfnamefont {Philip~Daniel}\
  \bibnamefont {Blocher}}, \bibinfo {author} {\bibfnamefont {Serwan}\
  \bibnamefont {Asaad}}, \bibinfo {author} {\bibfnamefont {Vincent}\
  \bibnamefont {Mourik}}, \bibinfo {author} {\bibfnamefont {Mark A.~I.}\
  \bibnamefont {Johnson}}, \bibinfo {author} {\bibfnamefont {Andrea}\
  \bibnamefont {Morello}}, \ and\ \bibinfo {author} {\bibfnamefont {Klaus}\
  \bibnamefont {M\o{}lmer}},\ }\href@noop {} {\enquote {\bibinfo {title}
  {Measuring out-of-time-ordered correlation functions without reversing time
  evolution},}\ } (\bibinfo {year} {2020}),\ \Eprint
  {http://arxiv.org/abs/2003.03980} {arXiv:2003.03980 [quant-ph]} \BibitemShut
  {NoStop}%
\bibitem [{\citenamefont {Heinzen}\ and\ \citenamefont
  {Wineland}(1990)}]{PhysRevA.42.2977}%
  \BibitemOpen
  \bibfield  {author} {\bibinfo {author} {\bibfnamefont {D.~J.}\ \bibnamefont
  {Heinzen}}\ and\ \bibinfo {author} {\bibfnamefont {D.~J.}\ \bibnamefont
  {Wineland}},\ }\bibfield  {title} {\enquote {\bibinfo {title}
  {Quantum-limited cooling and detection of radio-frequency oscillations by
  laser-cooled ions},}\ }\href {\doibase 10.1103/PhysRevA.42.2977} {\bibfield
  {journal} {\bibinfo  {journal} {Phys. Rev. A}\ }\textbf {\bibinfo {volume}
  {42}},\ \bibinfo {pages} {2977--2994} (\bibinfo {year} {1990})}\BibitemShut
  {NoStop}%
\bibitem [{\citenamefont {Monroe}\ \emph {et~al.}(1995)\citenamefont {Monroe},
  \citenamefont {Meekhof}, \citenamefont {King}, \citenamefont {Jefferts},
  \citenamefont {Itano}, \citenamefont {Wineland},\ and\ \citenamefont
  {Gould}}]{PhysRevLett.75.4011}%
  \BibitemOpen
  \bibfield  {author} {\bibinfo {author} {\bibfnamefont {C.}~\bibnamefont
  {Monroe}}, \bibinfo {author} {\bibfnamefont {D.~M.}\ \bibnamefont {Meekhof}},
  \bibinfo {author} {\bibfnamefont {B.~E.}\ \bibnamefont {King}}, \bibinfo
  {author} {\bibfnamefont {S.~R.}\ \bibnamefont {Jefferts}}, \bibinfo {author}
  {\bibfnamefont {W.~M.}\ \bibnamefont {Itano}}, \bibinfo {author}
  {\bibfnamefont {D.~J.}\ \bibnamefont {Wineland}}, \ and\ \bibinfo {author}
  {\bibfnamefont {P.}~\bibnamefont {Gould}},\ }\bibfield  {title} {\enquote
  {\bibinfo {title} {{Resolved-Sideband Raman Cooling of a Bound Atom to the 3D
  Zero-Point Energy}},}\ }\href {\doibase 10.1103/PhysRevLett.75.4011}
  {\bibfield  {journal} {\bibinfo  {journal} {Phys. Rev. Lett.}\ }\textbf
  {\bibinfo {volume} {75}},\ \bibinfo {pages} {4011--4014} (\bibinfo {year}
  {1995})}\BibitemShut {NoStop}%
\bibitem [{\citenamefont {Gardiner}\ \emph {et~al.}(1997)\citenamefont
  {Gardiner}, \citenamefont {Cirac},\ and\ \citenamefont
  {Zoller}}]{PhysRevA.55.1683}%
  \BibitemOpen
  \bibfield  {author} {\bibinfo {author} {\bibfnamefont {S.~A.}\ \bibnamefont
  {Gardiner}}, \bibinfo {author} {\bibfnamefont {J.~I.}\ \bibnamefont {Cirac}},
  \ and\ \bibinfo {author} {\bibfnamefont {P.}~\bibnamefont {Zoller}},\
  }\bibfield  {title} {\enquote {\bibinfo {title} {Nonclassical states and
  measurement of general motional observables of a trapped ion},}\ }\href
  {\doibase 10.1103/PhysRevA.55.1683} {\bibfield  {journal} {\bibinfo
  {journal} {Phys. Rev. A}\ }\textbf {\bibinfo {volume} {55}},\ \bibinfo
  {pages} {1683--1694} (\bibinfo {year} {1997})}\BibitemShut {NoStop}%
\bibitem [{\citenamefont {Kneer}\ and\ \citenamefont
  {Law}(1998)}]{PhysRevA.57.2096}%
  \BibitemOpen
  \bibfield  {author} {\bibinfo {author} {\bibfnamefont {B.}~\bibnamefont
  {Kneer}}\ and\ \bibinfo {author} {\bibfnamefont {C.~K.}\ \bibnamefont
  {Law}},\ }\bibfield  {title} {\enquote {\bibinfo {title} {Preparation of
  arbitrary entangled quantum states of a trapped ion},}\ }\href {\doibase
  10.1103/PhysRevA.57.2096} {\bibfield  {journal} {\bibinfo  {journal} {Phys.
  Rev. A}\ }\textbf {\bibinfo {volume} {57}},\ \bibinfo {pages} {2096--2104}
  (\bibinfo {year} {1998})}\BibitemShut {NoStop}%
\bibitem [{\citenamefont {Ben-Kish}\ \emph {et~al.}(2003)\citenamefont
  {Ben-Kish}, \citenamefont {DeMarco}, \citenamefont {Meyer}, \citenamefont
  {Rowe}, \citenamefont {Britton}, \citenamefont {Itano}, \citenamefont
  {Jelenkovi\ifmmode~\acute{c}\else \'{c}\fi{}}, \citenamefont {Langer},
  \citenamefont {Leibfried}, \citenamefont {Rosenband},\ and\ \citenamefont
  {Wineland}}]{PhysRevLett.90.037902}%
  \BibitemOpen
  \bibfield  {author} {\bibinfo {author} {\bibfnamefont {A.}~\bibnamefont
  {Ben-Kish}}, \bibinfo {author} {\bibfnamefont {B.}~\bibnamefont {DeMarco}},
  \bibinfo {author} {\bibfnamefont {V.}~\bibnamefont {Meyer}}, \bibinfo
  {author} {\bibfnamefont {M.}~\bibnamefont {Rowe}}, \bibinfo {author}
  {\bibfnamefont {J.}~\bibnamefont {Britton}}, \bibinfo {author} {\bibfnamefont
  {W.~M.}\ \bibnamefont {Itano}}, \bibinfo {author} {\bibfnamefont {B.~M.}\
  \bibnamefont {Jelenkovi\ifmmode~\acute{c}\else \'{c}\fi{}}}, \bibinfo
  {author} {\bibfnamefont {C.}~\bibnamefont {Langer}}, \bibinfo {author}
  {\bibfnamefont {D.}~\bibnamefont {Leibfried}}, \bibinfo {author}
  {\bibfnamefont {T.}~\bibnamefont {Rosenband}}, \ and\ \bibinfo {author}
  {\bibfnamefont {D.~J.}\ \bibnamefont {Wineland}},\ }\bibfield  {title}
  {\enquote {\bibinfo {title} {{Experimental Demonstration of a Technique to
  Generate Arbitrary Quantum Superposition States of a Harmonically Bound
  Spin-$1/2$ Particle}},}\ }\href {\doibase 10.1103/PhysRevLett.90.037902}
  {\bibfield  {journal} {\bibinfo  {journal} {Phys. Rev. Lett.}\ }\textbf
  {\bibinfo {volume} {90}},\ \bibinfo {pages} {037902} (\bibinfo {year}
  {2003})}\BibitemShut {NoStop}%
\bibitem [{\citenamefont {Rangan}\ \emph {et~al.}(2004)\citenamefont {Rangan},
  \citenamefont {Bloch}, \citenamefont {Monroe},\ and\ \citenamefont
  {Bucksbaum}}]{PhysRevLett.92.113004}%
  \BibitemOpen
  \bibfield  {author} {\bibinfo {author} {\bibfnamefont {C.}~\bibnamefont
  {Rangan}}, \bibinfo {author} {\bibfnamefont {A.~M.}\ \bibnamefont {Bloch}},
  \bibinfo {author} {\bibfnamefont {C.}~\bibnamefont {Monroe}}, \ and\ \bibinfo
  {author} {\bibfnamefont {P.~H.}\ \bibnamefont {Bucksbaum}},\ }\bibfield
  {title} {\enquote {\bibinfo {title} {{Control of Trapped-Ion Quantum States
  with Optical Pulses}},}\ }\href {\doibase 10.1103/PhysRevLett.92.113004}
  {\bibfield  {journal} {\bibinfo  {journal} {Phys. Rev. Lett.}\ }\textbf
  {\bibinfo {volume} {92}},\ \bibinfo {pages} {113004} (\bibinfo {year}
  {2004})}\BibitemShut {NoStop}%
\bibitem [{\citenamefont {Z\"ahringer}\ \emph {et~al.}(2010)\citenamefont
  {Z\"ahringer}, \citenamefont {Kirchmair}, \citenamefont {Gerritsma},
  \citenamefont {Solano}, \citenamefont {Blatt},\ and\ \citenamefont
  {Roos}}]{PhysRevLett.104.100503}%
  \BibitemOpen
  \bibfield  {author} {\bibinfo {author} {\bibfnamefont {F.}~\bibnamefont
  {Z\"ahringer}}, \bibinfo {author} {\bibfnamefont {G.}~\bibnamefont
  {Kirchmair}}, \bibinfo {author} {\bibfnamefont {R.}~\bibnamefont
  {Gerritsma}}, \bibinfo {author} {\bibfnamefont {E.}~\bibnamefont {Solano}},
  \bibinfo {author} {\bibfnamefont {R.}~\bibnamefont {Blatt}}, \ and\ \bibinfo
  {author} {\bibfnamefont {C.~F.}\ \bibnamefont {Roos}},\ }\bibfield  {title}
  {\enquote {\bibinfo {title} {Realization of a quantum walk with one and two
  trapped ions},}\ }\href {\doibase 10.1103/PhysRevLett.104.100503} {\bibfield
  {journal} {\bibinfo  {journal} {Phys. Rev. Lett.}\ }\textbf {\bibinfo
  {volume} {104}},\ \bibinfo {pages} {100503} (\bibinfo {year}
  {2010})}\BibitemShut {NoStop}%
\bibitem [{\citenamefont {Wu}\ \emph {et~al.}(2019)\citenamefont {Wu},
  \citenamefont {Zhang}, \citenamefont {Xie}, \citenamefont {Ou}, \citenamefont
  {Chen}, \citenamefont {Wu},\ and\ \citenamefont
  {Chen}}]{PhysRevA.100.062111}%
  \BibitemOpen
  \bibfield  {author} {\bibinfo {author} {\bibfnamefont {Chun-Wang}\
  \bibnamefont {Wu}}, \bibinfo {author} {\bibfnamefont {Jie}\ \bibnamefont
  {Zhang}}, \bibinfo {author} {\bibfnamefont {Yi}~\bibnamefont {Xie}}, \bibinfo
  {author} {\bibfnamefont {Bao-Quan}\ \bibnamefont {Ou}}, \bibinfo {author}
  {\bibfnamefont {Ting}\ \bibnamefont {Chen}}, \bibinfo {author} {\bibfnamefont
  {Wei}\ \bibnamefont {Wu}}, \ and\ \bibinfo {author} {\bibfnamefont
  {Ping-Xing}\ \bibnamefont {Chen}},\ }\bibfield  {title} {\enquote {\bibinfo
  {title} {Scheme and experimental demonstration of fully atomic weak-value
  amplification},}\ }\href {\doibase 10.1103/PhysRevA.100.062111} {\bibfield
  {journal} {\bibinfo  {journal} {Phys. Rev. A}\ }\textbf {\bibinfo {volume}
  {100}},\ \bibinfo {pages} {062111} (\bibinfo {year} {2019})}\BibitemShut
  {NoStop}%
\bibitem [{\citenamefont {Barton}(1986)}]{BARTON1986322}%
  \BibitemOpen
  \bibfield  {author} {\bibinfo {author} {\bibfnamefont {G.}~\bibnamefont
  {Barton}},\ }\bibfield  {title} {\enquote {\bibinfo {title} {Quantum
  mechanics of the inverted oscillator potential},}\ }\href {\doibase
  https://doi.org/10.1016/0003-4916(86)90142-9} {\bibfield  {journal} {\bibinfo
   {journal} {Annals of Physics}\ }\textbf {\bibinfo {volume} {166}},\ \bibinfo
  {pages} {322--363} (\bibinfo {year} {1986})}\BibitemShut {NoStop}%
\bibitem [{\citenamefont {Meekhof}\ \emph {et~al.}(1996)\citenamefont
  {Meekhof}, \citenamefont {Monroe}, \citenamefont {King}, \citenamefont
  {Itano},\ and\ \citenamefont {Wineland}}]{PhysRevLett.76.1796}%
  \BibitemOpen
  \bibfield  {author} {\bibinfo {author} {\bibfnamefont {D.~M.}\ \bibnamefont
  {Meekhof}}, \bibinfo {author} {\bibfnamefont {C.}~\bibnamefont {Monroe}},
  \bibinfo {author} {\bibfnamefont {B.~E.}\ \bibnamefont {King}}, \bibinfo
  {author} {\bibfnamefont {W.~M.}\ \bibnamefont {Itano}}, \ and\ \bibinfo
  {author} {\bibfnamefont {D.~J.}\ \bibnamefont {Wineland}},\ }\bibfield
  {title} {\enquote {\bibinfo {title} {{Generation of Nonclassical Motional
  States of a Trapped Atom}},}\ }\href {\doibase 10.1103/PhysRevLett.76.1796}
  {\bibfield  {journal} {\bibinfo  {journal} {Phys. Rev. Lett.}\ }\textbf
  {\bibinfo {volume} {76}},\ \bibinfo {pages} {1796--1799} (\bibinfo {year}
  {1996})}\BibitemShut {NoStop}%
\bibitem [{\citenamefont {Braunstein}\ and\ \citenamefont {van
  Loock}(2005)}]{Loock}%
  \BibitemOpen
  \bibfield  {author} {\bibinfo {author} {\bibfnamefont {Samuel~L.}\
  \bibnamefont {Braunstein}}\ and\ \bibinfo {author} {\bibfnamefont {Peter}\
  \bibnamefont {van Loock}},\ }\bibfield  {title} {\enquote {\bibinfo {title}
  {Quantum information with continuous variables},}\ }\href {\doibase
  10.1103/RevModPhys.77.513} {\bibfield  {journal} {\bibinfo  {journal} {Rev.
  Mod. Phys.}\ }\textbf {\bibinfo {volume} {77}},\ \bibinfo {pages} {513--577}
  (\bibinfo {year} {2005})}\BibitemShut {NoStop}%
\bibitem [{\citenamefont {Wang}\ \emph {et~al.}(2007)\citenamefont {Wang},
  \citenamefont {Hiroshima}, \citenamefont {Tomita},\ and\ \citenamefont
  {Hayashi}}]{WANG20071}%
  \BibitemOpen
  \bibfield  {author} {\bibinfo {author} {\bibfnamefont {Xiang-Bin}\
  \bibnamefont {Wang}}, \bibinfo {author} {\bibfnamefont {Tohya}\ \bibnamefont
  {Hiroshima}}, \bibinfo {author} {\bibfnamefont {Akihisa}\ \bibnamefont
  {Tomita}}, \ and\ \bibinfo {author} {\bibfnamefont {Masahito}\ \bibnamefont
  {Hayashi}},\ }\bibfield  {title} {\enquote {\bibinfo {title} {{{Quantum
  information with Gaussian states}}},}\ }\href {\doibase
  https://doi.org/10.1016/j.physrep.2007.04.005} {\bibfield  {journal}
  {\bibinfo  {journal} {Physics Reports}\ }\textbf {\bibinfo {volume} {448}},\
  \bibinfo {pages} {1--111} (\bibinfo {year} {2007})}\BibitemShut {NoStop}%
\bibitem [{\citenamefont {Hegde}\ \emph {et~al.}(2019)\citenamefont {Hegde},
  \citenamefont {Subramanyan}, \citenamefont {Bradlyn},\ and\ \citenamefont
  {Vishveshwara}}]{PhysRevLett.123.156802}%
  \BibitemOpen
  \bibfield  {author} {\bibinfo {author} {\bibfnamefont {Suraj~S.}\
  \bibnamefont {Hegde}}, \bibinfo {author} {\bibfnamefont {Varsha}\
  \bibnamefont {Subramanyan}}, \bibinfo {author} {\bibfnamefont {Barry}\
  \bibnamefont {Bradlyn}}, \ and\ \bibinfo {author} {\bibfnamefont {Smitha}\
  \bibnamefont {Vishveshwara}},\ }\bibfield  {title} {\enquote {\bibinfo
  {title} {{Quasinormal Modes and the Hawking-Unruh Effect in Quantum Hall
  Systems: Lessons from Black Hole Phenomena}},}\ }\href {\doibase
  10.1103/PhysRevLett.123.156802} {\bibfield  {journal} {\bibinfo  {journal}
  {Phys. Rev. Lett.}\ }\textbf {\bibinfo {volume} {123}},\ \bibinfo {pages}
  {156802} (\bibinfo {year} {2019})}\BibitemShut {NoStop}%
\bibitem [{\citenamefont {Morita}(2020)}]{Morita:2018sen}%
  \BibitemOpen
  \bibfield  {author} {\bibinfo {author} {\bibfnamefont {Takeshi}\ \bibnamefont
  {Morita}},\ }\bibfield  {title} {\enquote {\bibinfo {title} {{{Bound on
  Lyapunov exponent in $c=1$ matrix model}}},}\ }\href {\doibase
  10.1140/epjc/s10052-020-7879-9} {\bibfield  {journal} {\bibinfo  {journal}
  {Eur. Phys. J. C}\ }\textbf {\bibinfo {volume} {80}},\ \bibinfo {pages} {331}
  (\bibinfo {year} {2020})}\BibitemShut {NoStop}%
\bibitem [{\citenamefont {Ali}\ \emph {et~al.}(2020)\citenamefont {Ali},
  \citenamefont {Bhattacharyya}, \citenamefont {Haque}, \citenamefont {Kim},
  \citenamefont {Moynihan},\ and\ \citenamefont
  {Murugan}}]{PhysRevD.101.026021}%
  \BibitemOpen
  \bibfield  {author} {\bibinfo {author} {\bibfnamefont {Tibra}\ \bibnamefont
  {Ali}}, \bibinfo {author} {\bibfnamefont {Arpan}\ \bibnamefont
  {Bhattacharyya}}, \bibinfo {author} {\bibfnamefont {S.~Shajidul}\
  \bibnamefont {Haque}}, \bibinfo {author} {\bibfnamefont {Eugene~H.}\
  \bibnamefont {Kim}}, \bibinfo {author} {\bibfnamefont {Nathan}\ \bibnamefont
  {Moynihan}}, \ and\ \bibinfo {author} {\bibfnamefont {Jeff}\ \bibnamefont
  {Murugan}},\ }\bibfield  {title} {\enquote {\bibinfo {title} {Chaos and
  complexity in quantum mechanics},}\ }\href {\doibase
  10.1103/PhysRevD.101.026021} {\bibfield  {journal} {\bibinfo  {journal}
  {Phys. Rev. D}\ }\textbf {\bibinfo {volume} {101}},\ \bibinfo {pages}
  {026021} (\bibinfo {year} {2020})}\BibitemShut {NoStop}%
\end{thebibliography}%

\end{document}